\documentclass[12pt]{article}
\usepackage{amsmath,caption,indentfirst,graphicx,hyperref,amsfonts}
\usepackage{amssymb}

\usepackage[left=2cm,right=2cm,top=2cm,bottom=2cm,bindingoffset=0cm]{geometry}

\DeclareCaptionLabelSeparator{dot}{. }
\usepackage{xcolor}

\newcommand{\gD}{{\bf D}}
\newcommand{\gM}{{\bf M}}
\newcommand{\gF}{{\bf F}}

\newcommand{\gH}{{\rm{\bf H}}}
\newcommand{\gR}{{\rm{\bf R}}}

\newcommand{\gA}{{\rm{\bf A}}}

\newcommand{\ptl}{\partial}

\newcommand{\rU}{{\rm U}}
\newcommand{\rW}{{\rm W}}
\newcommand{\rA}{{\rm A}}
\newcommand{\rB}{{\rm B}}

\renewcommand{\Im}{{\rm Im}}
\renewcommand{\Re}{{\rm Re}}

\usepackage{placeins}    

\author{A.V. Shanin \and A.I. Korolkov \and K.S. Kniazeva}
\title{Saddle point method for transient processes in waveguides}
\date{}

\begin{document}

\maketitle


\begin{abstract}

A modification of the saddle point method is proposed for computation of non-stationary wave processes (pulses)
in waveguides. The dispersion diagram of the waveguide is continued analytically. A set of possible saddle points 
on the dispersion diagram is introduced. A method of checking whether the particular saddle points 
contribute terms to the field decomposition is proposed. A classification of the waveguides based on the topology of the set of possible saddle points 
is outlined. 

\end{abstract}


%

\section{Introduction}
\label{sec1}

Non-stationary wave processes (short pulses) in a lossless closed waveguide of regular cross-section are studied.
The waveguide can have an arbitrary nature, e.g.\ acoustical, electromagnetic, or elastic one. A waveguide is assumed to be excited by a point source localized in time. We prefer not to specify mathematically what is a ``regular cross-section'', 
but on the physical level the absence of the {\em black holes\/} \cite{Mironov2020} is implied.  
The aim of the paper is to describe the wave processes in such a waveguide asymptotically, taking the observation point 
far enough from the source point.  

For simplicity, we describe the waveguide by the so-called WaveFEM (also called WFEM or waveguide-FEM) method \cite{Aalami1973,Finnveden2008}. 
The waveguide is assumed to be discrete in the transverse direction and continuous in the longitudinal direction. 
Moreover, we study very simple waveguides  having only 2 or 5 nodes on the cross-section. On the one hand, such  
waveguides exhibit a nontrivial behavior themselves. On the other hand, the techniques of analysis remain similar when the amount of nodes per a cross--section is large (say, several thousand, which is enough for rather complicated waveguides).     

A formal solution for such a problem can be easily obtained with the help of the Fourier transform. 
As the result, one obtains a double integral representation of the field. After application of the residue theorem, 
the solution becomes expressed as a sum of several oscillating integrals with integration contours laying on a branched 
Riemann surface. 
We develop a technique to evaluate such integrals asymptotically.

Let us give a review of existing methods. For definiteness, asymptotics of integral of the form
\begin{equation}
\label{OscIntegral}
I(x,V) =\int_{-\infty}^{\infty}f(k)\exp\{ i\, x \, g(k,V) \} \, dk
\end{equation}
are looked for. Functions $f(k)$ and $g(k)$ are supposed to be branching analytic functions, $V$ is some real parameter (the 
ratio $x/t$ in the waveguide context, $x$ is the coordinate of the observation point, and $t$ is the time), 
and $g(k)$ is the phase function, supposed to be real-valued on the real axis. 
The coordinate $x$ is large, and we are building the asymptotics as $x \to \infty$.
The easiest way to estimate such an integral is to apply stationary phase method.  The idea behind the method is quite simple: the vicinities of the points  $k_j$ where the phase function $g(k)$ is stationary ($dg/dk=0$) provide the principal 
contributions to the integral. These contributions are found with the help of integration by parts.
According to Erdelyi \cite{Erdelyi1955} the stationary phase method was outlined by Cauchy, and was first used by Stokes \cite{Stokes2009} to study Airy function. The general formulation was given by Lord Kelvin \cite{Thomson1887}.

More precise asymptotic estimation of the integral (\ref{OscIntegral}) can be obtained with the help of the saddle point method also known as steepest descent method. The general idea is to deform the contour of integration in such a way that the imaginary part of phase function $g(k)$ increases as fast as possible.  The contour will pass through the saddle points $k_j$ where $dg/dk=0$ (as above), but $k_j$ is not necessary real. Contributions calculated in vicinities of such points provide 
an asymptotic estimation to the integral. 

Particularly, let the function $g(k,V)$ be expanded as the Taylor series  near $k_j$:
\begin{equation}
\label{phase_series}
g(k,V)\approx g(k_j,V) + \frac{(k-k_j)^2}{2} \frac{d^2g}{dk^2}(k_j,V),
\end{equation}
and $f(k)$ is approximated by its value in the saddle point $f(k_j)$. Then the integral (\ref{OscIntegral}) is evaluated as the Poisson's integral.
This procedure was introduced by Riemann \cite{Riemann1953} and developed by Debye~\cite{Debye1909}. One can expect the resulting approximation to be better than the stationary phase one, since it takes into account contributions from a wider set of saddle points. However, the problem of building the exact steepest descent contour is very expensive in terms of computational efforts. Fortunately, there is no need to know the whole contour, one just need to know which saddle points contribute to the integral and which do not. Recently, this issue was addressed in the context of quantum field theory \cite{Witten2010}. Particularly, in \cite{Feldbrugge2017,Feldbrugge2019} an algorithm to deform the original contour into a sum steepest descent contours was introduced. 
The authors introduced a vector field indicating the growth and decay directions of the imaginary part of the phase function $g(k,V)$. Then, using the Picard--Lefschetz analysis \cite{Arnold2012} they proved that the contour deformation along such decay directions (they are called inward flows in the paper) leads to the steepest descent contour. This contour is refereed to as the Lefschetz' thimble.

In the current paper we develop a similar approach, but move it further.  We introduce a simple criterion to determine whether the particular saddle points are active (i.e.\ contribute to the integral) or not.
Then, we study the family of integrals (\ref{OscIntegral}) 
indexed by  the parameter $V$.  A set of possible saddle points is introduced for this family. This set is called 
{\em the carcass of the dispersion diagram}. 
Then the points of the carcass are classified. 
In the context of the waveguide problems each active point of the carcass corresponds to some physical pulse. We claim that the carcass 
can be used as a tool for analysis and classification of waveguides. 
Note that we do not restrict ourselves to meromorphic functions in (\ref{OscIntegral}) as the authors of \cite{Feldbrugge2017,Feldbrugge2019} did, and this makes a substantial difference in the analysis.

The theory of resurgent functions was used to study the steepest descent contours and corresponding asymptotics of integrals.
This theory was introduced by Ecalle \cite{Ecalle1981}, and was developed later by \cite{Sternin1997,Delabaere1999}. Unfortunately, it seems to be very abstract and hard to use.

There is a great number of works devoted to the cases where the saddle point method fails. Let us mention some of them. There are two general situations when the saddle point method cannot be applied directly:  the saddle point is close to other saddle points; there are two saddle points on a single steepest descent contour.

The first case leads to invalidity of the expansion (\ref{phase_series}). Thus, the 
phase function $g(k)$ has a critical point of a higher order. Suprisingly, the classification of such points can be done in the context of real functions. Particularly, one may study the phase function as a function of real variables $k$ and $V$. The latter allows one to solve the problem using the catastrophe theory \cite{Poston2014}. Thom \cite{Thom2018} introduced seven singularities (also called degenerate critical points), and named them as the seven catastrophes: fold, cusp, swallowtail, butterfly, and  elliptic, hyperbolic and parabolic umbilic. The behavior of corresponding oscillating integrals was later studied by Arnold \cite{Arnold2012}.

The second  case is known as the Stokes phenomenon \cite{Olver1990}. It is explained by the fact that the saddle points form a discrete set, and some of them may leave or hit the steepest descend contour as the parameter $V$ changes. So, for a critical value of parameters the saddle point contour will go through two saddle points. This situation is unstable and a subtle change of parameters will destroy it.  The well known example of such a behavior is provided by Airy function:
\begin{equation}
{\rm Ai}(\xi) = \frac{1}{2\pi i}\int_{\gamma}\exp\left\{i\tau^3 /3-i\xi \tau \right\}d\tau ,
\end{equation}
where the contour $\gamma$ comes from infinity with the argument equal to $-\pi/3$ and goes to infinity with the argument 
equal to~$\pi/3$. The exponential has two saddle points. 
The steepest descent contours can pass through both of them or only one~\cite{Wright1980}. 
Berry~\cite{Berry1988} introduced a jump-free asymptotics (a hyperasymptotics)
to deal with the discontinuity of the saddle point asymptotics.

The structure of the paper is as follows. In section~\ref{subsec21} we introduce the WaveFEM equation and obtain a formal solution in terms of rapidly oscillating integrals.  In section~\ref{subsec23} we obtain the stationary phase asymptotics. A numerical example is presented showing how inaccurate this asymptotic can be even for the WaveFEM equations of dimension~2.  
In section~\ref{sec3} we develop the saddle point asymptotic. A theorem that the original contour of integration can be deformed into a sum of steepest descent contours is proven. The concept of carcass of a dispersion diagram is introduced. The saddle point classification is explained. In section~\ref{sec6} we provide  numerical examples of application of the developed method to  different WaveFEM systems.


\section{Problem statement. WaveFEM equations}
\label{subsec21}

The problem of pulse propagation in a waveguide will be described by the  WaveFEM
equation. Such an equation may be considered as
a finite-dimension numerical approximation of a continuous waveguide, or, alternatively, as
an exact description of a waveguide with a discrete cross-section.
We stress that the most natural way to obtain the WaveFEM equation for a given waveguide is to apply the finite element method 
(FEM) to the cross-section of the waveguide. Indeed, the denser grid is taken, the more exact results are obtained.  

The waveguide is described by a vector $\rU (t, x)$ of dimension $J \times 1$.
The parameter $J$ is the number of degrees of freedom, by which a single
cross-section of the waveguide is described.
The variables $t$ and $x$ are the time and the longitudinal coordinate.
We assume that this vector obeys
the WaveFEM equation of the form
\begin{equation}
\left(
\gD_2  \ptl_x^2 + \gD_1 \ptl_x + \gD_0 - \gM \ptl_t^2
\right) \rU(t , x) =  \gF \, \delta (t) \, \delta (x).
\label{e2101}
\end{equation}
Hear $\gM$, $\gD_0$, $\gD_1$, $\gD_2$ are  real constant square matrix coefficients
of dimension $J \times J$.
The right-hand side is the excitation of the waveguide. The
constant vector $\gF$ of dimension
$J \times 1$ is the transversal profile of the excitation. One can see that
the excitation is localized in time and space.
We are looking for a solution of (\ref{e2101}) obeying an additional requirement:
the solution $\rU (t, x)$ should be causal,
i.~e.\
\begin{equation}
\rU(t, x) = 0 \quad \mbox{for} \quad t < 0.
\label{e2102}
\end{equation}

We assume that the matrices $\gM$, $\gD_0$, $\gD_1$, $\gD_2$ possess the following properties \cite{Finnveden_2004}:
\begin{equation}
\gM^T = \gM,
\qquad
\gD_0^T = \gD_0,
\qquad
\gD_1^T = -\gD_1,
\qquad
\gD_2^T = \gD_2.
\label{e2103}
\end{equation}
Besides, we assume that $\gM$ and $\gD_2$ are positively defined matrices, and
$-\gD_0$ is a non-negatively defined matrix.
These properties  can be easily established if the equation
(\ref{e2101}) is obtained from the finite element formulation.

A simplest waveguide of the type (\ref{e2101}) is a {\em scalar waveguide\/} 
with $J = 1$, i.e.\ all coefficient matrices are scalar. According to (\ref{e2103}),
$\gD_1 = 0$ in this case. One can see that the scalar WaveFEM equation is 
a Klein--Gordon equation. This equation describes a wave process possessing a cut-off frequency.  

The waveguide modes can be introduced as follows. Let be
$\rU (t, x) = \rU \exp\{ i k x - i \omega t\}$ for some complex parameters $(\omega , k)$,
which are the temporal circular frequency and the longitudinal wavenumber.
Such modes should obey a homogeneous version of (\ref{e2101}). As the result, 
for each fixed $k$ one
gets a generalized eigenvalue problem
\begin{equation}
\left(
- k^2 \gD_2 + ik \gD_1 + \gD_0
\right) \rU = - \omega^2 \gM \, \rU,
\label{e2104}
\end{equation}
for which $\omega^2$ is an eigenvalue and $\rU$
is an eigenvector.

For each $k$ the eigenvalue problem (\ref{e2104}) has, generally, $J$ eigenvalues
$\omega_j^2 (k)$, $j = 1, \dots , J$. Respectively, generally,
there are $2J$ values $\pm \omega_j (k)$.

The waveguide is assumed to have no energy loss or gain. We formulate a consequence of this
as the {\em reality statement} \cite{Finnveden_2004}: If $k$ is real then all values $\omega_j (k)$ are real.
The validity of this statement can be understood as follows. For real $k$, one can convert the waveguide
into a resonator by taking a segment $0 \le x \le 2\pi / k$ and imposing the periodicity
condition on its ends. The values $\omega_j (k)$ are then the eigenfrequencies
of the resonator. Due to the absence of energy loss or gain, these frequencies should be real.

In all examples below we take for simplicity 
\begin{equation}
\gD_1 = 0.
\label{e2105}
\end{equation}
Under this condition, the reality statement is valid provided $\gD_0$, $\gD_2$, and $\gM$
obey the conditions (\ref{e2103}). If $\gD_1 \neq 0$ the condition of validity of the
reality statement is more complicated.

\vskip 6pt
{\bf Remark.}
The WaveFEM model (\ref{e2101}) is universal and convenient, being compared, say,
with the description continuous in the transversal dimensions.
This model can be obtained by discretization of the cross-section for virtually
any waveguide of acoustic, electromagnetic, or elastic nature.
The model is finite-dimensional, and the integral representations of the form
(\ref{e2201}) are obtained in a universal way.
We should admit, however, that the WaveFEM model (\ref{e2101})
seems to be not very efficient for describing ray processes in the near-field zone.

\subsection{Integral representations of the transient field}
\label{subsec22}

Let find a causal solution of the equation (\ref{e2101}). Introduce the Fourier transform with respect to $x$, and the Laplace transform with respect to $t$:
\begin{equation}
\tilde w(\omega,k) = \frac{1}{4\pi^2}\int\limits_{-\infty}^{\infty}\int\limits_{0}^{\infty}w(t,x)\exp\{-i k x + i\omega t\}dt \, dx. 
\end{equation}
Note that the Laplace variable is chosen as $-i\omega$, so the resulting notations are ``Fourier-like''. The inverse transform is
\begin{equation}
w(t,x) = \int\limits_{-\infty}^{\infty}\int\limits_{-\infty+i\varepsilon}^{\infty+i\varepsilon}\tilde w(\omega,k)\exp\{i k x - i\omega t\}d\omega \, dk,
\end{equation}
where $\varepsilon$ is an arbitrary positive parameter.
%
Applying these transformations to (\ref{e2101}), we obtain the double integral representation of the wavefield:
\begin{equation}
\rU(t , x) = \frac{1}{4 \pi^2}
\int \limits_{-\infty }^{\infty }
\int \limits_{-\infty + i \varepsilon}^{\infty + i \varepsilon}
\frac{\gA(\omega , k)\, \gF}{D(\omega , k)} e^{i k x - i \omega t}
d\omega \, dk,
\label{e2201}
\end{equation}
where $D(\omega , k)$ is the {\em dispersion function}:
\begin{equation}
D(\omega , k) = {\rm det} \left(
-k^2 \gD_2 + i k \gD_1 + \gD_0 + \omega^2 \gM
\right),
\label{e2202}
\end{equation}
and $\gA(\omega , k)$ is the adjugate matrix of
$-k^2 \gD_2 + i k \gD_1 + \gD_0 + \omega^2 \gM$:
\begin{equation}
\left(
-k^2 \gD_2 + i k \gD_1 + \gD_0 + \omega^2 \gM
\right)^{-1} = \frac{\gA (\omega , k)}{D(\omega , k)}.
\label{e2203}
\end{equation}
Note that elements of $\gA (\omega , k)$ are polynomials
of $\omega$ and~$k$.

The internal integral in (\ref{e2201}) can be taken by the residue method.
Note that $D(\omega , k)$ is a polynomial with respect to $\omega$ and~$k$.
The roots of this polynomial with respect to $\omega$, i.~e.\
the roots of the {\em dispersion equation}
\begin{equation}
D(\omega , k) = 0
\label{e2204}
\end{equation}
for fixed $k$, are the values $\pm \omega_j (k)$, which are square roots of 
eigenvalues of the problem (\ref{e2104}). Since $k$ is real in the domain of integration, the roots $\pm \omega_j (k)$ are real (see the {\em reality statement}).
For $t > 0$ we can close the contour of the internal integral in the lower half-plane.
The integral becomes a sum of residual terms:
\[
\rU (t , x) = \frac{1}{2\pi i} \sum_{j = 1}^J
\int \limits_{- \infty}^{\infty}
\left[
\frac{\gA(\omega_j (k)  , k)\, \gF}{\ptl_\omega D(\omega_j (k) , k)}
\exp \{i k x - i \omega_j (k) t \}
-
\right.
\]
\begin{equation}
\left.
\frac{\gA(\omega_j (k)  , k)\, \gF}{\ptl_\omega D(\omega_j (k) , k)}
\exp \{i k x + i \omega_j (k) t \}
\right]
dk
\label{e2205}
\end{equation}

Our aim is to estimate the integral (\ref{e2205})
for relatively large $x$ and~$t$.

\section{Estimation of  (\ref{e2205}) by the stationary phase method}
\label{subsec23}

The stationary phase method is the simplest way to estimate
the integral (\ref{e2205}).
This method works well for very large $t$ and~$x$ (\cite{Erdelyi1955}).

One can see that (\ref{e2205}) is a sum of $2J$ integrals of the
form
\begin{equation}
\rW(t, x) = \int \limits_{-\infty}^{\infty}
\rB( k) \exp \{ i k x - i \varpi (k) t \} \, dk
\label{e2301}
\end{equation}
where $\varpi (k)$ is one of the functions $\pm \omega_j (k)$,
\[
\rB(k) = \frac{1}{2\pi i}
\frac{\gA(\varpi (k) , k) \gF}{\ptl_\omega D(\varpi (k) , k)}.
\]
Note that (\ref{e2205}) is a sum of all such integrals, i.e.\ the integration is held over 
all sheets of a multivalued function.  

Let us estimate a single integral (\ref{e2301}).
According to the reality statement, the function $\varpi (k)$ is real.
Introduce the group velocity
\begin{equation}
v_{\rm gr} (k) = \frac{d \varpi (k)}{dk}.
\label{e2302}
\end{equation}
This group velocity is also a real function for real~$k$.

Introduce the {\em formal velocity\/}
\begin{equation}
V \equiv x / t.
\label{e2303}
\end{equation}
We are estimating $\rW(t, x)$ for fixed $V$ and for $x \to \infty$.
Find all real points $k_{*m}$, for which
\begin{equation}
v_{\rm gr} (k_{*m})  = V.
\label{e2304}
\end{equation}
These points are the stationary points of the phase function $g(k) = g(k, V)$:
\[
g(k) \equiv k - \varpi (k) / V ,
\qquad 
k x - \varpi (k) \, t  = x g(k).
\]

Index $m$ in $k_{*m}$ takes some integer values. Positions of the points $k_{*m}$ and
their amount depend on~$V$: $k_{*m} = k_{*m}(V)$.
The integral (\ref{e2301}) is  estimated as a sum of  contributions provided by the stationary phase points:
\begin{equation}
\rW \approx
\sum_m \exp\{ - {\rm sign}(\alpha_{*m}) \pi i / 4\}
\exp\{ i(k_{*m} x - i \varpi(k_{*m}) t) \}
\sqrt{\frac{2 \pi}{|\alpha_{*m} | t}}
\rB( k_{*m}),
\label{e2305}
\end{equation}
\begin{equation}
\alpha_{*m} \equiv \left. \frac{d^2 \varpi}{dk^2} \right| _{k_{*m}}.
\label{e2306}
\end{equation}

\begin{figure}
\centering
\includegraphics[scale = 0.8]{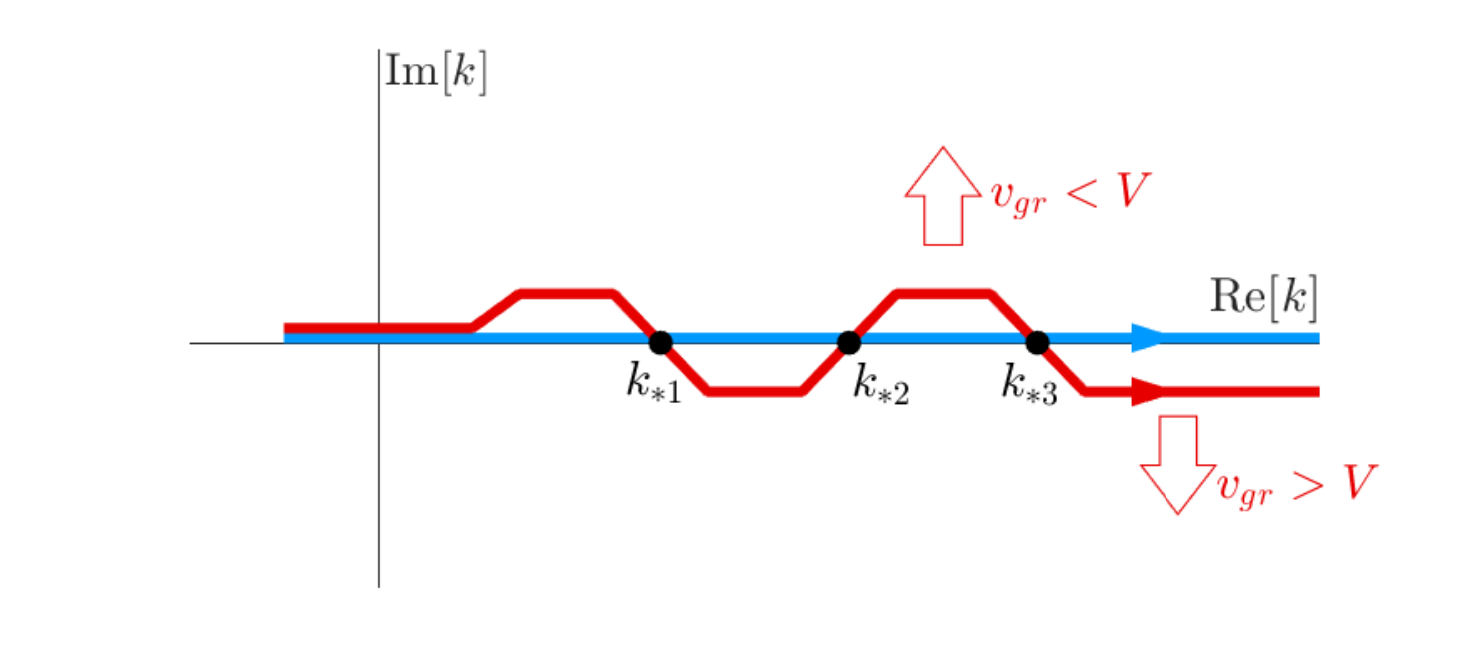}
\caption{Contour deformation for the stationary phase method}
\label{StPhM}
\end{figure}

\vskip 6pt
\noindent
{\bf Remark.} We would like to to emphasize the connection between the stationary phase method and the saddle point method. Such a connection was noticed by Poincaré \cite{Poincare1904} and Copson \cite{Copson1946}. Namely, the stationary phase points are also the saddle points of the analytically continued phase function. The integration contour can be deformed from the real axis of $k$ into the complex domain of~$k$ in order to get an exponential decay of the integrand 
almost everywhere (see~\figurename~\ref{StPhM}).  The deformation is rather simple:
the fragments with $V > v_{\rm gr}(k)$ are shifted into the upper half-plane, the fragments $V < v_{\rm gr}(k)$ are shifted into the lower half-plane, the points with
$k_{*m}$ remain on the real axis. All the shifts are assumed to be small, i.~e.\ occur within some narrow strip ${\rm Im}[k] < \Delta k$.

One can see that such deformation makes ${\rm Im}[g(k)] \ge 0$ everywhere on the contour.
Moreover, ${\rm Im}[g(k)] > 0$ everywhere except the saddle points $k_{*m}$,
where ${\rm Im}[g(k)] = 0$. For $x$ large enough, all parts of the
integration contour except the neighborhoods of the saddle points
$k_{*m}$ yield exponentially small contributions to the integral, and the saddle points contributions lead to~(\ref{e2305}).

Note that 
the terms described by (\ref{e2305}) possess a power decay, but they are not exponentially decaying.

Following \cite{Borovikov1994}, one can introduce the domain of influence (DOI) for each point
$k_{*m}$ as a circle with the centre at $k_{*m}$ and the radius of
\begin{equation}
\Delta k = \sqrt{\frac{2}{|\alpha_{*m}| t}} = \sqrt{\frac{2 V}{|\alpha_{*m}| x}}
\label{e2307}
\end{equation}
indicating the size of the area over which the saddle point integral
is actually taken. Using the concept of DOI, one can estimate the validity
of the stationary phase technique. Namely, the DOIs of different saddle points should not overlap,
and the function $\rB$ should be approximately constant within each DOI.

The stationary phase method is rather simple, and it works well
for very large~$x$ (and fixed~$V$). However, for the values of $x$ that are not very large,
it may happen that there exist some field components that are exponentially small, but still
cannot be ignored. Typically, such field components are described by saddle points
terms with the saddle points located not on the real axis.

To take into account all saddle points, one should, instead,  apply a saddle point method,
in which one should deform the initial integration contours into sums of steepest descent contours.
The rest of the paper describes this process.

An important example of an exponentially decaying wave component
is the forerunner in a waveguide \cite{Shanin2017}. Despite of decay, it may be faster
than all non-decaying components in some domain of $t$, $x$ and in some frequency band,
so it cannot be ignored in applications.

\subsection{Motivating numerical example. Stationary phase vs.\ saddle point method}
\label{subsec24}

Here we propose a simple demonstration of influence of non-real saddle point.
Consider the WaveFEM equation of the form (\ref{e2101}) of dimension $J =2$ with
\begin{equation}
\gM = {\rm I},
\quad
\gD_2 = \left( \begin{array}{cc}
4 & 0 \\
0 & 1
\end{array} \right),
\quad
\gD_1 = 0 ,
\quad
\gD_0 = \left( \begin{array}{cc}
-1 & -2 \\
-2 & -5
\end{array} \right),
\quad
\gF = \left( \begin{array}{c}
1 \\
0
\end{array} \right).
\label{e2401}
\end{equation}
One can see that the matrices $\gM$ and $\gD_2$ are diagonal, while $\gD_0$
is not. Thus, one can consider the equation (\ref{e2101}), (\ref{e2401}) as a
system of two interacting scalar waveguides.

\begin{figure}[h!]
\centering
\includegraphics[width = 8cm]{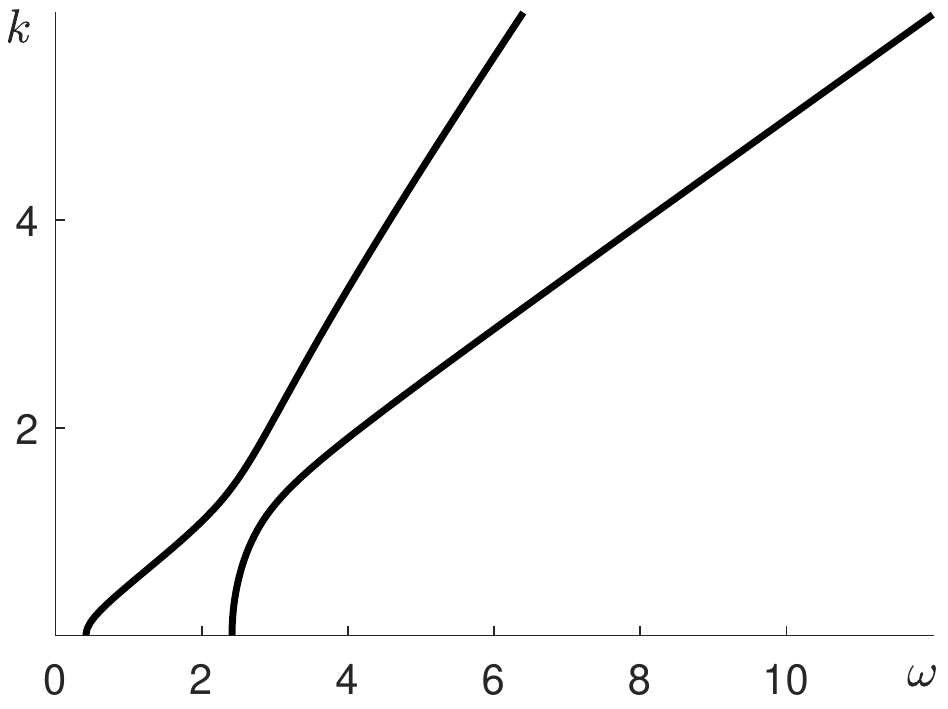}
\hspace{4ex}
\includegraphics[width = 8cm]{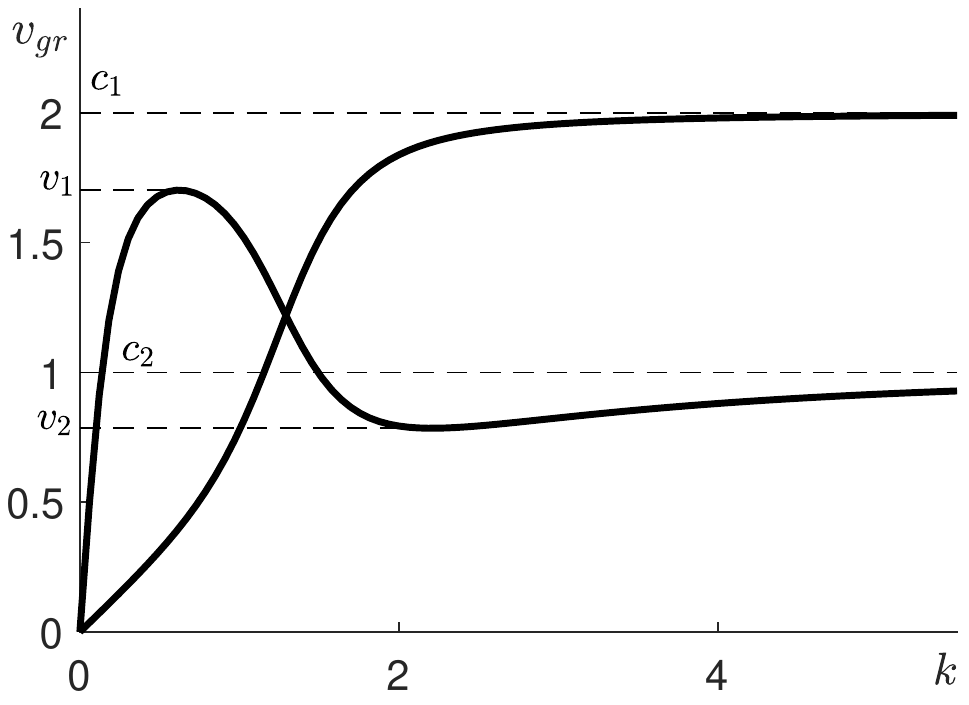}
\caption{Dispersion diagram (left) and group velocities (right) for the waveguide with coefficients (\ref{e2401})}
\label{fig01}
\end{figure}

One can easily build the dispersion diagram for this waveguide
(see \figurename~\ref{fig01}, left)
and compute the
group velocities
(see \figurename~\ref{fig01}, right).

According to the graph of $v_{\rm gr} (k)$, the limiting group velocities
as $|k| \to \infty$ are $c_1  = 2$ and $c_2 = 1$.
These values are, indeed, the square roots of the eigenvalues of the limiting
generalized eigenvalue problem following from (\ref{e2104}):
\begin{equation}
\gD_2 \rU= c^2 \gM \rU.
\label{e2402}
\end{equation}
An advanced study of the equation (\ref{e2101}) shows that the higher velocity
of these two, namely
$c_1$, is the limiting velocity in the system, i.~e.\
\begin{equation}
\rU (t, x)  =0
\qquad
\mbox{for}
\qquad
 |x| > c_1 t.
\label{e2403}
\end{equation}


Then, there are critical group velocities $v_1 = 1.702$ and $v_2 = 0.7848$
corresponding to the local maximum and the local minimum of corresponding branch
of the graph. Thus, for the stationary phase method, there are 4 regimes:

\vskip 6 pt
\begin{tabular}{ll}
$V > c_1$       & no points $k_{*m}$, \\
$v_1 < V <c_1$  & one point $k_{*m}$ on the half-axis $k \in (0 , \infty)$ \\
$V_2 < V <v_1$  & three points $k_{*m}$ on the half-axis $k \in (0 , \infty)$ \\
$v_2 < V <c_2$  & four points $k_{*m}$ on the half-axis $k \in (0 , \infty)$ \\
$0 < V <v_2$  & two points $k_{*m}$ on the half-axis $k \in (0 , \infty)$ \\
\end{tabular}
\vskip 6 pt

Indeed, for each stationary phase point $k_{*m}$ there exists a symmetrical point $-k_{*m}$
on the half-axis $(-\infty , 0)$.

The stationary phase method should lose its validity near the points
$V = c_1, c_2, v_1, v_2$, where DOIs of some stationary points overlap.

\begin{figure}[h!]
\centering
\includegraphics[width = 8cm]{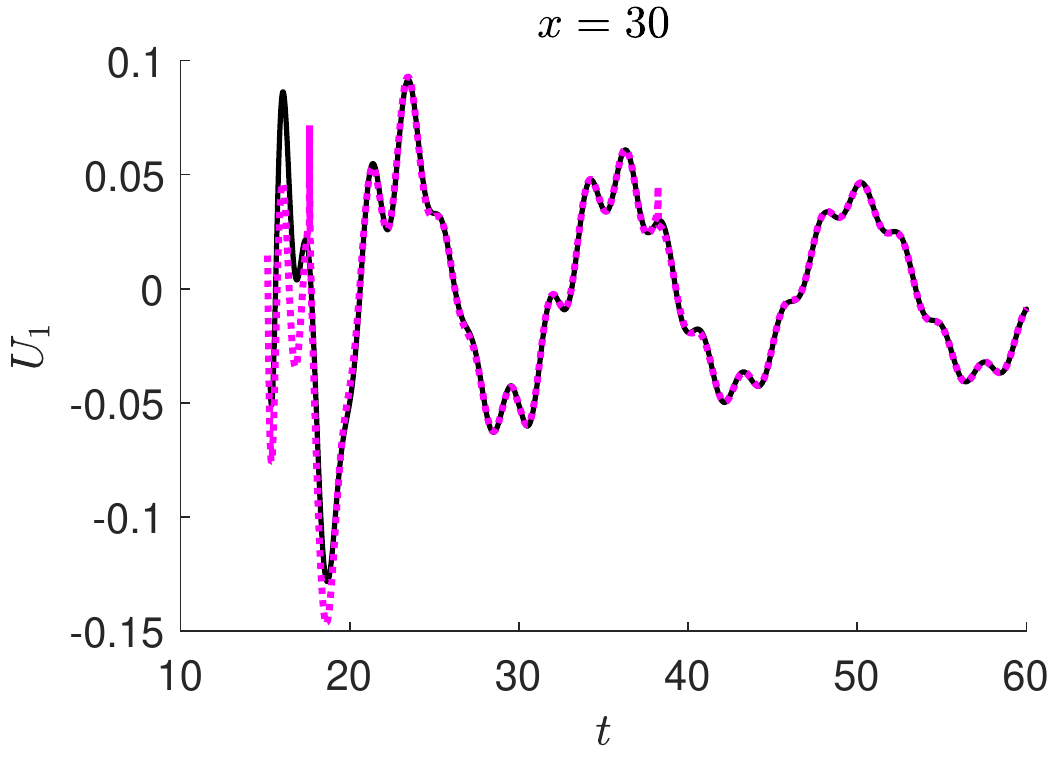}
\includegraphics[width = 8cm]{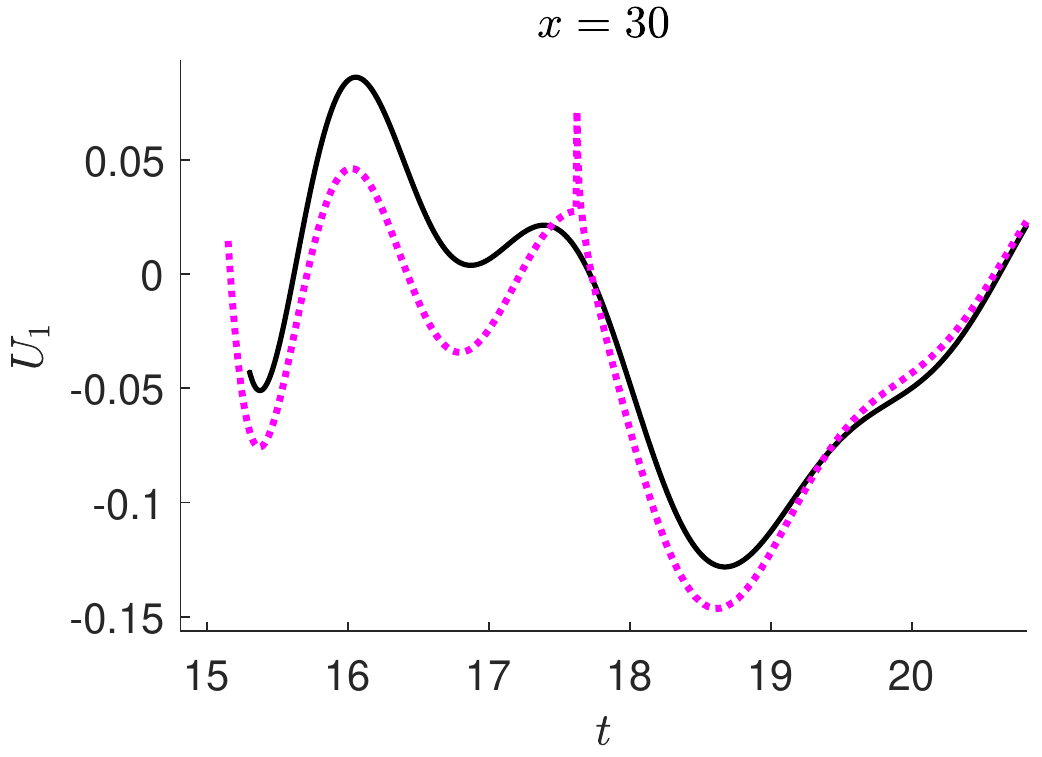}
\caption{The component $U_1$ computed by taking the integral (\ref{e2205}) (black line) and by
the stationary phase method (magenta dots). The right part of the figure shows the zoomed version of the left figure}
\label{fig02}
\end{figure}

In \figurename~\ref{fig02}, left,
we compare the integral (\ref{e2205}) with the
result of application of the stationary phase method computed by the formula
(\ref{e2305}). The value of $x$ is taken equal to 30. The component $U_1$ of $\rU$ is shown.
The graph shows the function $U_1 (t,x)$ for fixed $x$ as a function of~$t$.
Black line corresponds to the integral (\ref{e2205}) computed numerically, and the magenta dots correspond to
the stationary phase method. One can see that for $t > X/v_2$ the signal is approximately a sum of two sinusoidal components.
This is because there exist two values of $k_{*m}$ (and, respectively, two values of $\omega$)
providing these components.

One can see that near the time moments $t_1 = x/v_1$ and $t_2 = x/v_2$ the stationary phase
method becomes invalid. One can also see that the stationary phase method
yields poor accuracy for $x/ c_1 < t <x / v_1$. A zoomed version of the
same figure is shown in
\figurename~\ref{fig02}, right.

\begin{figure}[h!]
\centering
\includegraphics[width = 8cm]{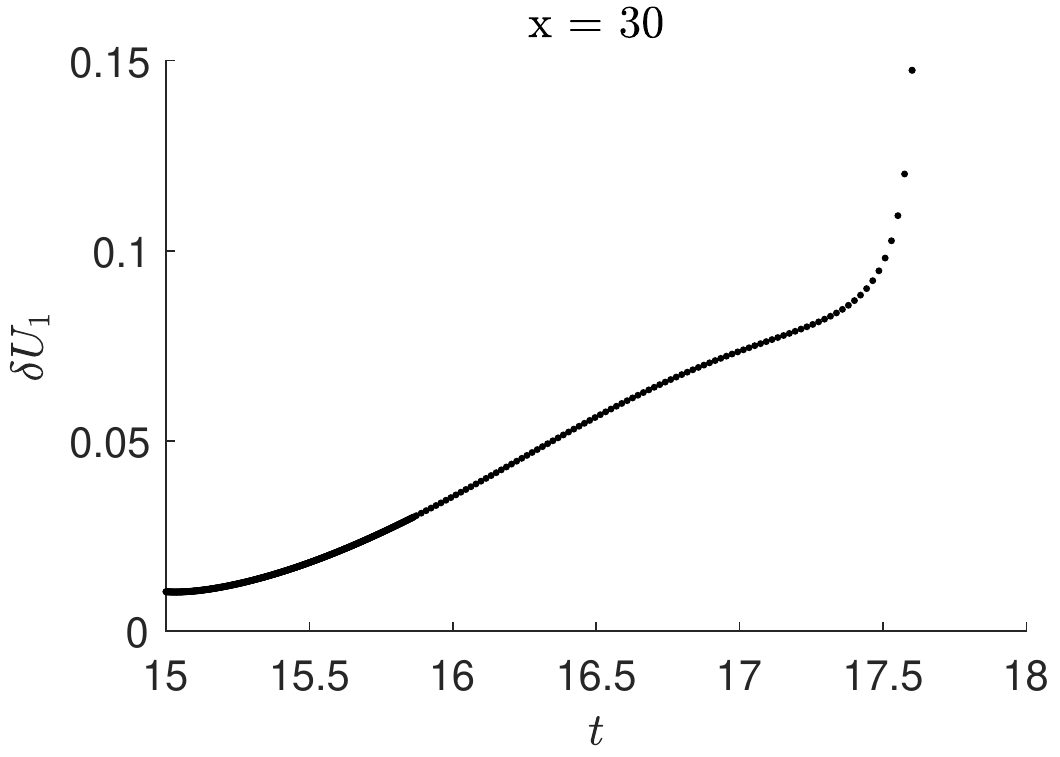}
\includegraphics[width = 8cm]{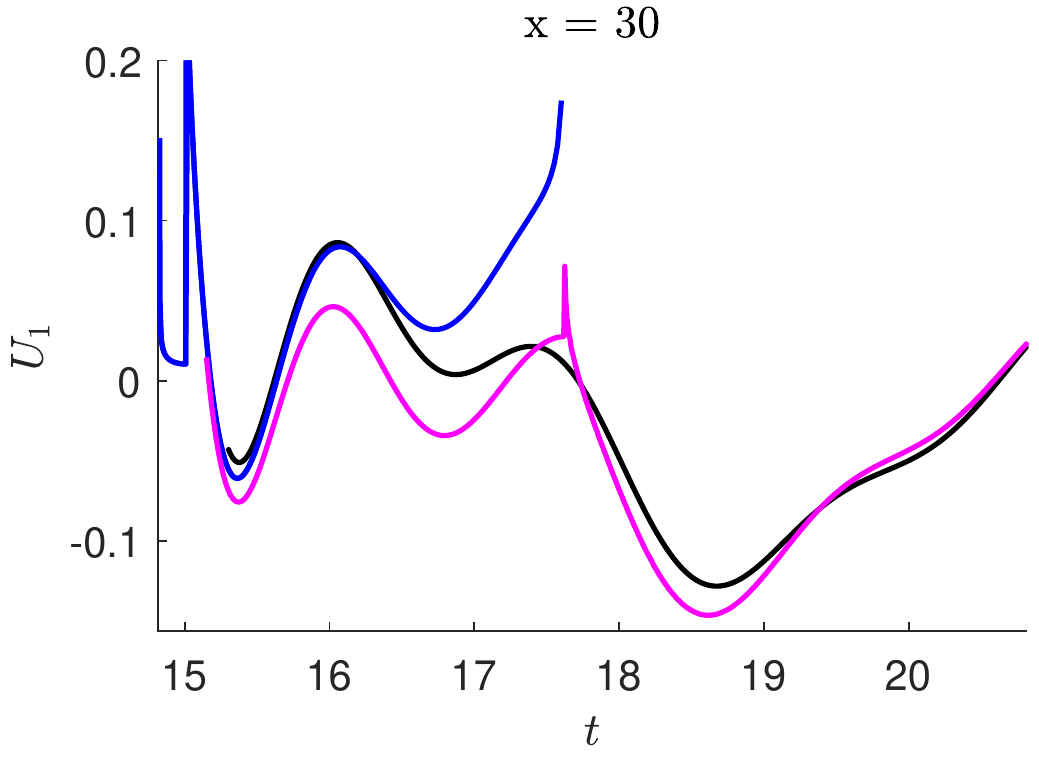}
\caption{Left: the correction $\delta U_1$ caused by some complex saddle points. Right: component $U_1$ obtained by numerical calculation of (\ref{e2205}) (black),
the stationary phase method (magenta), and the saddle point method (blue)}
\label{fig03}
\end{figure}

The reason of such a discrepancy is the presence of some other saddle points $k_{*m}$
with non-zero imaginary parts that should be taken into account.
Such saddle points are described below in details. In \figurename~\ref{fig03}, left
we demonstrate the wave component corresponding to these complex saddle points alone.
In \figurename~\ref{fig03}, right, we show the  saddle point asymptotic vs.\ the exact signal.
The black line denotes the exact solution (\ref{e2205}), the magenta line corresponds to the stationary phase method, and the blue line is obtained by the saddle point method. One can see that the accuracy of the saddle point method is increased in a certain domain comparatively to that of the stationary phase method.

In a slightly speculative way, we claim that each saddle point corresponds to a certain physical process in a 
waveguide, so including a complex saddle point into consideration may be important from the physical point of view.


\section{Solution by the saddle point method}
\label{sec3}

\subsection{The dispersion diagram as a complex manifold}

Let us make two important mathematical preparatory steps concerning the
representation (\ref{e2205}).
The {\em first step\/} is similar to that of \cite{Shanin2017,Shanin2018}.
Let the values $\pm \omega_j (k)$ be different values of a branched function
$\omega (k)$ of the complex variable~$k$.
Denote the Riemann surface of this function by~$\gR$. This surface
has $2J$ sheets.

Redefine the function $\rB (k)$ as
\begin{equation}
\rB (k) = \frac{1}{2\pi i}
\frac{\rA (\omega (k) , k) \gF}{\ptl_\omega D(\omega(k) , k)}
\label{e3101}
\end{equation}
Since the functions $\rA (\omega , k)$ and $\ptl_\omega D(\omega , k)$
are polynomials of $\omega$ and $k$, the function $\rB (k)$ is possibly branched on $\mathbb{C}$,
and it is single-valued on~$\gR$.

One can see that (\ref{e2205}) can be rewritten as
\begin{equation}
\rU(t, x) = \sum_{n  = 1}^{2J}
\int_{\gamma_n} \rB(k) \exp \{ i k x - i \omega(k) t \} \, dk =
\int_{\gamma_1 + \dots \gamma_{2J}} \rB(k) \exp \{ i k x - i \omega(k) t \} \, dk
,
\label{e3102}
\end{equation}
where $\gamma_j$ are the samples of the real $k$-axis on~$\gR$.
All contours $\gamma_j$ are passed in the positive direction of~$k$.
One can see that (\ref{e3102}) is the integral of a single-valued function on
$\gR$ taken over a complicated contour (a sum of $2J$ simple contours).
In what follows we are going to keep the ansatz (\ref{e3102}), and deform the
integration contours into the steepest descend contours.

The representation (\ref{e3102}) itself is enough to obtain the results that are presented below. 
However, it has a considerable fault
decreasing its universality. Namely, $\gR$ has branch points that should be taken into account in
the contour deformation process. Moreover, the denominator of (\ref{e3101})
should have zeros at the branch points of $\gR$, and this should be also important for estimation of the integral (\ref{e3101}) 
in its initial form.

To overcome this difficulty, we perform the {\em second step\/}.
Namely, consider all points $(\omega , k) \in \mathbb{C}^2$ obeying the dispersion
equation (\ref{e2204}). Denote the set of all such points by~$\gH$.
Generally, $\ptl_\omega D$, $\ptl_k D$,  and $D$
are not equal to zero simultaneously (see the remark below).
Thus all points of $\gH$ are regular (in the sense of classification of points of surfaces defined implicitly)
and $\gH$ is a smooth 2D surface in a 4D space. Indeed,
$\gR$ is a projection of $\gH$ onto the $k$-plane.
Denote this projection by~$\psi$. This projection works as
\begin{equation}
\psi: \quad (\omega(k) , k) \to k.
\label{e3103}
\end{equation}
The preimages (with respect to $\psi$)
of the branch points of $\gR$ are regular points of  $\gH$, and the
branch points are just ``folds'' caused by the projection.

The surface $\gH$ is the {\em dispersion diagram\/} of the waveguide.

The surface $\gH$ can be considered as a {\em complex manifold} \cite{Shabat1992},
i.e.\ a complex structure can be defined on~$\gH$. Such a structure is
defined as follows. The surface $\gH$ is split into neighborhoods small enough, and
a local complex variable is introduced in each neighborhood, describing the neighborhood in a trivial way. The formulae of transition
between the local variables should be biholomorphic in intersections of neighborhoods.

One can consider $k$, or $\omega$, or other variable as a local variable. 
A good choice of local variable is $k$ for all neighborhoods not including the branch points 
of $\gR$, and $\omega$ near the branch points. 

The existence of a complex structure on $\gH$ means that one can introduce a contour of integration
on~$\gH$. For such an integration, one should have an analytic differential 1-form on $\gH$
(this may be a restriction on $\gH$ of an analytical differential 1-form in $\mathbb{C}^2$)
and an oriented integration contour on~$\gH$.
The Cauchy theorem is valid for such an integration, i.~e.\ one can deform the integration contours on $\gH$ without changing the value of the integral provided the deformation
occurs in the domain of regularity of the form.

Consider the following differential 1-form in the space $\mathbb{C}^2$
of variables $(\omega , k)$:
\begin{equation}
\Phi =
\frac{1}{2 \pi i}
\rA(\omega , k) \gF \exp\{ {\it i k x - i \omega t }\} \Psi,
\label{e3104}
\end{equation}
where $\Psi$ is the differential 1-form
\begin{equation}
\Psi = \frac{dk}{\ptl_\omega D(\omega , k)}.
\label{e3105}
\end{equation}
Let $\Phi|_{\gH}$ and $\Psi|_{\gH}$ be the restrictions of the forms
$\Phi$ and $\Psi$ on~$\gH$.

Let $\gamma'_n$, $n = 1 , \dots , 2J$ be oriented contours on $\gH$, which are
preimages (with respect to $\psi$) of the oriented real $k$-axis,
i.~e.\ they are all continuous contours $(\omega(k) , k)$ with real $k$
changing from $-\infty$ to~$\infty$.
One can rewrite (\ref{e3102}) as follows:
\begin{equation}
\rU(t, x) = \int_{\gamma'_1 + \dots + \gamma'_{2J}} \Phi|_{\gH}
\label{e3106}
\end{equation}

Let us demonstrate that the form $\Phi|_{\gH}$ is, generally, regular.
The factor $\rA(\omega , k)$ is a polynomial of $\omega$ and~$k$, thus,
it is regular.
The form $\Psi |_{\gH}$ with $\Psi$ defined by (\ref{e3105})
is regular everywhere except the preimages of
the branch points of $\gR$, where $\ptl _{\omega}D = 0$.
According to the theorem about implicit functions,
\begin{equation}
\left. \frac{dk}{\ptl_\omega D(\omega , k)} \right|_{\gH}
= -
\left. \frac{d\omega}{\ptl_k D(\omega , k)} \right|_{\gH}.
\label{e3107}
\end{equation}
Thus, at the preimage of a branch point of $\gR$, one can take the second
representation of $\Psi |_{\gH}$. As we have already noted, we suppose that $\ptl_k D$ cannot be zero
at the preimages of the branch points of $\gR$, where $\ptl D_{\omega} = 0$, so the form is regular (see also remark below).

A summary of this section is as follows: below we study
the representation (\ref{e3106}). We assume that the surface $\gH$ is smooth everywhere
(although it may have a very sophisticated topology, see \cite{Shanin2018}),
and the differential form $\Phi |_\gH$ is regular everywhere, thus the integration contours can be deformed in a rather free way.

We realize that the formalism of complex manifold and analytical differential forms
may be not very popular among the specialists in waveguides. Thus, to avoid making the
whole contents below useless, we can simplify the
formulation of the statement above. The contour deformation can be
performed in a usual way on the Riemann surface $\gR$,
{\em and the branch points of $\gR$ require no special consideration}.
As one can see, the usage of the complex manifold concept
makes the situation simpler.

\vskip 6 pt
\noindent
{\bf Remark. }
As it is well-known, a point of $\gH$ is not regular if for some $(\omega , k)$ 
\[
D(\omega , k) =0,
\qquad 
\ptl_\omega D(\omega , k) = 0, \qquad  \ptl_k D(\omega , k) = 0.
\]
This forms a system of 3 complex
restrictions for 2 complex degrees of freedom. Generally, such a system is overdefined and  has no solutions.
Conversely, a system of two equations  $\ptl_\omega D = 0$, $\ptl_k D = 0$,
generally, has a discrete set of solutions (points). Such solutions are
introduced in \cite{ShestopalovShestopalov} as {\em critical points\/}
of the dispersion diagram. Such critical points do not belong to a dispersion
diagram $D= 0$, but, being located near such a diagram, can cause a
peculiar behavior of the latter. We assume that the case of Shestopalov's
critical point belonging to the dispersion diagram is quite special, and it falls
beyond the scope of the current paper.

\subsection{Saddle points on a complex manifold}
\label{subsec41}

Define a complex phase function $g$ on $\gH$:
\begin{equation}
\label{e4101}
g(p;V)\equiv k - \omega/V, \qquad p = (\omega,k) \in \gH.
\end{equation}
The exponential factor in (\ref{e3102}) reads as $\exp\{ i x g \}$.
Indeed, $g(p;V)$ is an analytic function on $\gH$.

Introduce also a real function $g_i$:
\begin{equation}
\label{e4102}
g_i(p;V) \equiv \Im[g(p;V)], \qquad p = (\omega,k) \in \gH.
\end{equation}
The exponential factor  is large if $g_i < 0$, and small if $g_i > 0$.

Define the saddle points and the steepest descend contours on $\gH$. Let us start with the saddle points. Consider the function $g_i(p;V)$ in some neighborhood of a point $p \in \gH$. Suppose this point has a local complex coordinate $z$. A saddle point $p \in \gH$ is a point, at which
\begin{equation}
\label{e4103}
\ptl_{z'}g_i(p;V) = 0, \qquad \partial_{z''}g_i(p;V) = 0,
\end{equation}
where $z' = \Re[z]$, $z'' = \Im[z]$, i.~e.~the gradient of $g_i$ with respect to $p$, is equal to zero. The point 
$p$ of~(\ref{e4103}) cannot be a local minimum or a local maximum of $g_i$ since $g_i$ is a harmonic function. Since there are two restrictions~(\ref{e4103}) on $\gH$, the set of saddle points is discrete.
For $\gH$ defined by~(\ref{e2204}), there is a finite set of saddle points, since this equation is algebraic. The saddle points are stable: a small change of $V$ cannot kill or create a saddle point, it can only slightly change positions of saddle points.

One can see that
\begin{equation}
\label{e4104}
\ptl_{z'}g_i(p;V) = \Im[\ptl_z g(p;V)],
\qquad
\ptl_{z''}g_i(p;V) = \Re[\ptl_z g(p;V)].
\end{equation}
Thus, $p$ is a saddle point if and only if the complex derivative $\ptl_zg$ is equal to zero. The complex derivative is not defined invariantly on $\gH$, i.~e.~the value of the derivative depends on the choice of the local complex coordinate. However, the fact that the complex derivative is equal to zero is invariant. Thus, one can check, say, condition
\begin{equation}
\label{e4105}
\ptl_{\omega}g(p;V) = 0
\end{equation}
or
\begin{equation}
\label{e4106}
\ptl_{k}g(p;V) = 0,
\end{equation}
taking $\omega$ or $k$ as the local variable, to establish that point $p$ is a saddle point. 

Note that equations~(\ref{e4105}) and~(\ref{e4106}) are equivalent to the equation
\begin{equation}
\label{e4107}
\frac{d\omega}{dk} = V.
\end{equation}
If, as usual, the group velocity is formally introduced as
\begin{equation}
\label{vgr}
\begin{gathered}
v_{\rm gr}(p) \equiv \frac{d \omega}{d k}, \qquad p \in \gH,
\end{gathered}
\end{equation}
Note that according to the theorem about the implicit function,
\begin{equation}
v_{\rm gr} = - \frac{\ptl_k D}{\ptl_\omega D}.
\label{vgr1}
\end{equation}
Then, (\ref{e4107}) reads as
\begin{equation}
\label{e4108}
v_{\rm gr}(p) = V.
\end{equation}

The saddle points will be marked with a star decoration. The notation $p_{\ast}(V)$ indicates the dependence of the saddle point $p_{\ast} \in \gH$ on the parameter $V$.

Now let us define the steepest descend contours. A direction of steepest descend of the exponential factor $\exp\{ixg(p;V)\}$ corresponds to the steepest growth of $g_i$. Thus, to build the required contours, it is necessary to define the vector field of gradient vector of $g_i$ on $\gH$:
\begin{equation}
\label{e4109}
\bigtriangledown g_i \equiv \begin{pmatrix} \ptl_{z'}g_i \\ \ptl_{z''}g_i \end{pmatrix}.
\end{equation}
 and integrate this vector field, i.~e.~to find smooth curves on $\gH$ that are tangential to the gradient of $g_i$ at each point. 
Note that (\ref{e4109}) defines the gradient only locally, since a change of local complex variable $z$ does not 
affect the direction of the gradient.  


An alternative way to define the steepest descent contours is to notice that the steepest descend contours are, at the same time, the stationary phase contours, i.~e.~the contours on which the value $\Re[g]$ is constant.

Below, the steepest descend contours are referred to as {\em streams}. We don't use the word ``contours'' (of steepest descends) to avoid confusion with the integration contours.

\begin{figure}
\centering
\includegraphics[scale = 1]{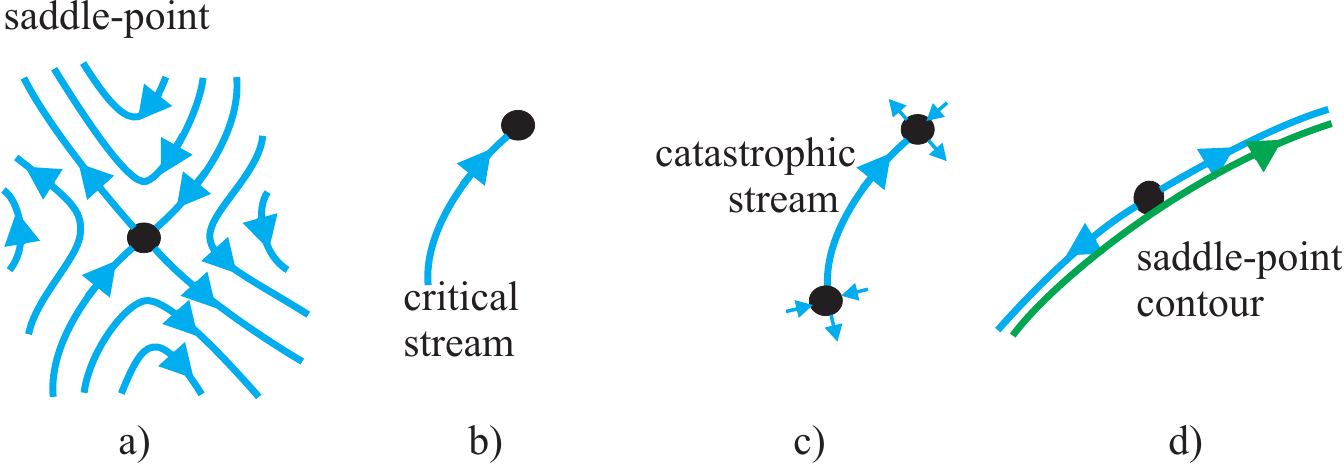}
\caption{Different streams}
\label{figB3101a}
\end{figure}

Streams are oriented in the direction of growth of $g_i$. A single stream can be emitted from any point of $\gH$ that is not a saddle point. From a saddle point, two streams can be emitted. Besides, two streams come to this point. A small neighborhood 
of a saddle point 
on $\gH$ is shown in \figurename~\ref{figB3101a},~a). Streams are blue lines with arrows, the saddle point is a circle.

If a stream hits a saddle point it is referred to as a {\em critical stream\/} (see \figurename~\ref{figB3101a},~b)). If a stream goes from one saddle point to another, it will be called a {\em catastrophic stream\/} (see \figurename~\ref{figB3101a},~c)). Existence of catastrophic streams is known as Stokes phenomenon \cite{Olver1990}, and corresponding values of the parameter $V$ form so-called Stokes set. Since there is a discrete set of saddle points, the existence of a catastrophic stream is an unstable event, i.e.\ almost any small change of parameters should destroy it. We call the values of $V$ for which catastrophic streams exist somewhere on $\gH$ the catastrophic values of~$V$.

Consider a saddle point and two streams going from it. Assume that the streams do not hit other saddle points. The streams form a contour going on $\gH$ from infinity to infinity through the saddle point. Let this contour be oriented somehow. The contour is called a saddle point contour. A saddle point contour is shown by the green line in \figurename~\ref{figB3101a},~d).


\subsection{The main statement of the saddle point method}
\label{subsec42}
Here we give a formulation of the main theorem of the saddle point method:

\vskip 6pt
\noindent
{\bf Theorem}
{ \label{theorem}
{\em For non-catastrophic values of $V$, any contour $\gamma_n'$ of (\ref{e3106}) on $\gH$ can be homotopically transformed into a sum of several saddle point contours. The transformation at infinity is eligible, i.e.\ the contour passes through the zones of decay of the exponential factor.}
\vskip 6 pt
}

\vskip 6pt
\noindent
{\bf Remark.} A similar theorem was proved in \cite{Feldbrugge2019} but only for meromorphic functions.

\vskip 6pt

The proof of the theorem is as follows. 
Take some contour $\gamma_n'$ from (\ref{e3106}). This is a single continuous preimage of the real axis of $k$ on $\gH$. Emit streams from each point of $\gamma_n'$ and study behavior of these streams.

Let there be several (possibly, zero) saddle points on $\gamma_n'$. Denote them by $p_{\ast 1}, \dots, p_{\ast m}$.

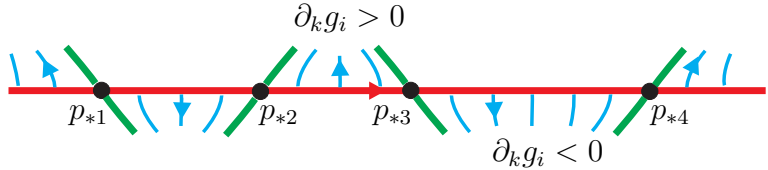
\begin{figure}[h]
\begin{center}
\begin{picture}(320,192)(0,0)
\put(16,16){
\includegraphics[width=288\unitlength]{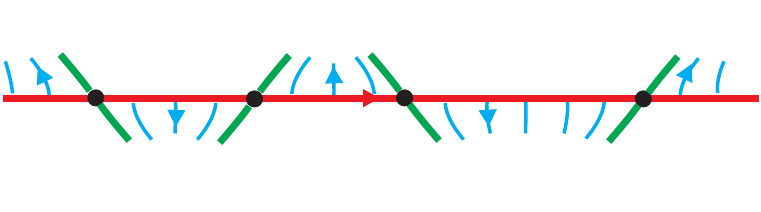}}

\put(200,30){
\makebox(0,0)[lb]{$\partial_k g_i < 0$}}
\put(125,81){
\makebox(0,0)[lb]{$\partial_k g_i > 0$}}
\put(40,46){
\makebox(0,0)[lb]{$p_{*1}$}}
\put(112,46){
\makebox(0,0)[lb]{$p_{*2}$}}
\put(155,46){
\makebox(0,0)[lb]{$p_{*3}$}}
\put(260,46){
\makebox(0,0)[lb]{$p_{*4}$}}
\end{picture}
\end{center}
\caption{Streams near $\gamma_n'$}
\label{fig4_2}
\end{figure}

The function $g$ is real and $g_i$ is equal to zero on $\gamma_n'$. Since $g_i$ grows on a stream, the streams cannot cross any preimage of real axis $k$. Everywhere on $\gamma_n'$ except the saddle points $p_{\ast j}$ the streams are normal to $\gamma_n'$. If $\ptl_k g_i > 0$ ($\ptl_k g_i < 0$) then the streams go to the upper (lower)
half-plane of $k$. In \figurename~\ref{fig4_2} we plot the behavior of the streams near $\gamma_n'$. Contour $\gamma_n'$ is red, the streams are blue, the saddle point contours passing through $p_{\ast j}$ are green. We do not draw orientation on the green contours at this moment.

\begin{figure}
\centering
\includegraphics[scale = 1]{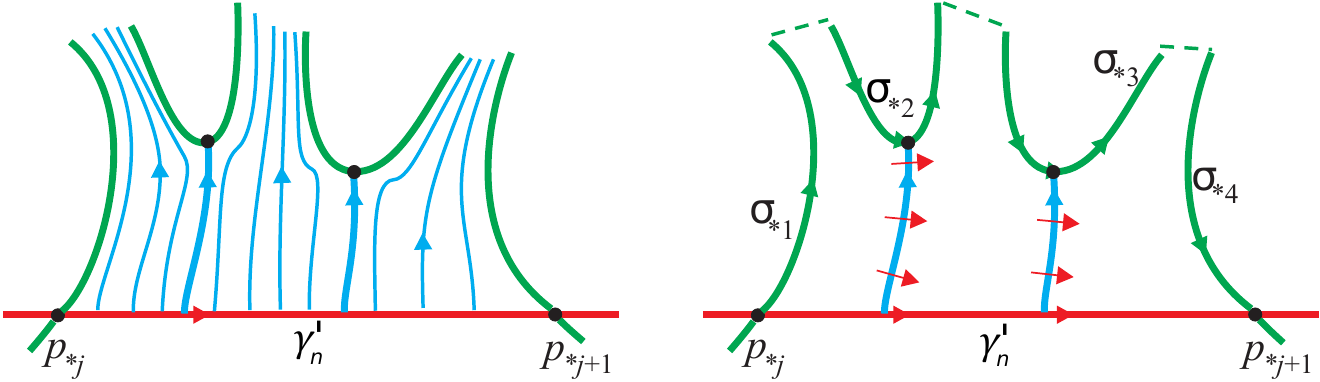}
\caption{Single-connected domain occupied by streams }
\label{fig4_3}
\end{figure}

Consider a segment of $\gamma_n'$ between some points $p_{\ast j}$ and $p_{\ast j+1}$ and continue the streams emitted from the points of this segment. The result is shown in \figurename~\ref{fig4_3}, left. Most of the streams will go to infinity, and a finite number of streams are critical, 
so they hit some saddle points outside $\gamma_n'$. Such critical streams are shown by bold blue.

When a critical stream hits a saddle point, there appear two other streams going from the saddle point and corresponding to growth of $g_i$. These streams are shown by green. Such streams form saddle point contours.

One can see that the domain occupied by streams emitted from the points of the segment $(p_{\ast j}, p_{\ast j+1})$ of $\gamma_n'$ is a single-connected domain by construction, although it is drawn on the manifold $\gH$, which may be not single-connected. This domain is bounded by the red line (the segment $(p_{\ast j}, p_{\ast j+1})$ of $\gamma_n'$), green lines (the parts of saddle point contours $\sigma_{\ast l}$), and some infinitely remote parts, where $g_i$ tends to infinity (\figurename~\ref{fig4_3}, right).

Let us assign direction to the saddle point contours. First consider the halves of the saddle point contours $\sigma_{\ast 1}$ and $\sigma_{\ast 4}$ in \figurename~\ref{fig4_3}, right. Assign the direction to these parts in a trivial way: 
contour $\sigma_{\ast 1}$ goes {\em from\/} the point $p_{\ast j}$, and contour $\sigma_{\ast 4}$ goes {\em to\/} the point $p_{\ast j+1}$. For the contours $\sigma_{\ast 2}$ and $\sigma_{\ast 3}$, the direction of these contours is ``propagated'' from the direction of $\gamma_n'$ via the critical streams. This process is explained by \figurename~\ref{fig4_3}, right. The direction ``propagates'' with thin red arrays from $\gamma_n'$ to the saddle point lines.

One can see that the oriented segment $(p_{\ast j},p_{\ast j+1})$ of $\gamma_n'$ can be homotopically deformed into the sum of oriented green contours
$\sigma_{\ast 1} + \sigma_{\ast 2} + \sigma_{\ast 3} + \sigma_{\ast 4}$, which are the saddle point contours or parts of them. There are also dashed green lines, but the integrand of (\ref{e3106}) is exponentially small on these parts of the contour, if $\gamma_n'$ contains no saddle points.

All parts of $\gamma_n'$ can be considered like this. A slightly modified argument works for half-infinite fragments or even for the whole contour. Summing up all deformed contours, one obtains the statement of the theorem.

\vskip 6pt

Let us discuss the theorem and its proof.
\begin{enumerate}
\item
The proof of the theorem gives a recipe of finding the saddle point contours, into which the contours $\gamma_n'$, 
$n = 1, \dots, 2J$ can be deformed. One should a) take the saddle point contours for the saddle points located on $\gamma_n'$, and b) emit streams from each point of $\gamma_n'$, find all streams hitting the saddle points, and take saddle point contours for all such saddle points. The proof also contains recommendations for finding the orientation of the saddle point contours.

\item
Let us consider the case when two critical streams hit the same saddle point (see \figurename~\ref{fig4_4}). The streams can be emitted from the same contour $\gamma_n'$ or from different contours $\gamma_{n_1}'$ and $\gamma_{n_2}'$. Let us prove that the resulting saddle point contours are oriented in the opposite way, and thus cancel each other.
\begin{figure}
\centering
\includegraphics[scale = 1]{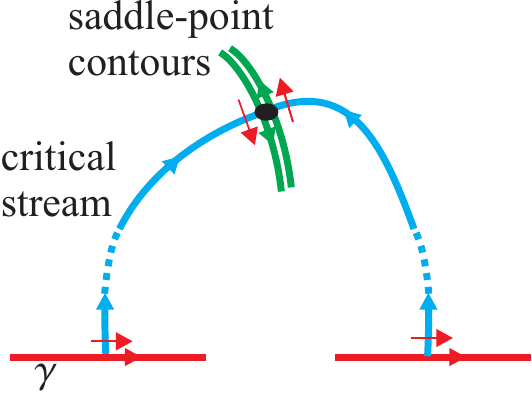}
\caption{Two critical streams hitting the same saddle point}
\label{fig4_4}
\end{figure}
The proof is based on the following statements. a) Since the streams cannot cross the preimages of the real $k$-axis, both critical streams are emitted from the real $k$-axis to the upper half-plane of $k$ or both to the lower-plane of $k$. Thus, the pair of red and blue arrows are oriented in the same way in terms of local variable $k$ at the emission points. b) On the complex manifold $\gH$ any change of local variables $(z_1',z_1'') \rightarrow (z_2', z_2'')$ between intersecting neighborhood keeps the orientation (i.~e.~the Jacobian of the transform is positive). To demonstrate this, write down the Jacobian and use the Cauchy-Riemann conditions for the function $z_2(z_1)$:
\begin{equation}
\label{VarTransCond}
\det \begin{pmatrix} \ptl z_2'/\ptl z_1' \quad \ptl z_2'/\ptl z_1'' \\
\ptl z_2''/\ptl z_1' \quad \ptl z_2''/\ptl z_1'' \end{pmatrix} = \left| \frac{d z_2}{d z_1} \right| ^2.
\end{equation}
Thus, in each neighborhood through which the stream passes, the mutual orientation of the red arrow the blue arrow is defined 
consistently and it does not change along the stream.  
c) At the meeting point, blue arrows are oriented in the opposite directions. Thus, the red
arrows are oriented in the opposite direction. The statement proven here will be used later as a criterion of an active saddle point.
\item
Consider the set of streams described in the proof of the theorem. Denote the domain occupied by the streams emitted from the contours $\gamma_n'$ by $\Xi$. Perform a small variation of the parameter $V$.  One can see that the structure of the saddle point contours changes only if some saddle point crosses the boundary of $\Xi$. This phenomena is known as Stokes transition \cite{Feldbrugge2019}. Namely, a saddle point can either enter $\Xi$ or exit $\Xi$. Anyway, a rebuilding of the system of saddle point contours happens when a saddle point contour from one saddle point hits another saddle point, i.e.\ at catastrophic values of $V$. An example of such a rebuilding is shown in \figurename~\ref{fig4_5}. The saddle point $p_{\ast}$ crosses one of the saddle point contours. Three states are shown. In part a) the point $p_{\ast}$ belongs to $\Xi$. In the part b) the catastrophic situation is shown. In part c) point $p_{\ast}$ is located outside $\Xi$.

\begin{figure}
\centering
\includegraphics[scale = 1]{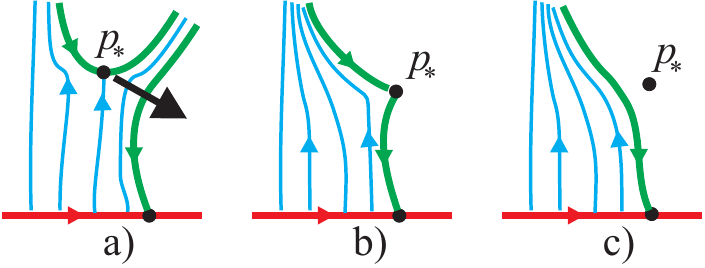}
\caption{Saddle point crossing the saddle point contour as $V$ changes}
\label{fig4_5}
\end{figure}
\item
Fix the value of $V$. Let it not be catastrophic. Let 
$p_{\ast j}=p_{\ast j}(V)=(\omega_{\ast j},k_{\ast})$, $j = 1, \dots, m$ be all saddle points on $\gH$. Let $\gamma_{\ast j}$ be corresponding saddle point contours oriented in an arbitrary way. The main theorem can be formulated as follows.
\vskip 6pt

{\em Each contour of integration $\gamma_{n}'$ of (\ref{e3106}) can be deformed into a sum of the saddle point contours with some coefficients $a_i$
\begin{equation}
\label{theorem2v}
\int \limits_{\gamma_1' + \dots + \gamma_{2J}'} \Phi |_{\gH}  = \sum_{j=1}^{m} a_j \int \limits_{\gamma_{\ast j}} \Phi |_{\gH} , \quad a_j \in [-1, 0, 1]
\end{equation}
by an admissible deformation.}
\vskip 6pt

\item
The integral over each $\gamma_{\ast j}$ in (\ref{theorem2v}) can be estimated by a standard saddle point method. The estimations is
\begin{equation}
\int \limits_{\gamma_{\ast j}} \Phi|_{\gH} \approx I_j x^{-1/2} \exp\{ix(k_{\ast j} - \omega_{\ast j} / V)\},
\end{equation}
where $(\omega_{\ast j}, k_{\ast j})$ is the saddle point corresponding to $\gamma_{\ast j}$,
\begin{equation}
I_j = \pm (2\pi)^{3/2} i^{1/2}V^{1/2} \left( \partial^2_k\omega (p_{\ast j}) \right) ^{-1/2},
\end{equation}
the sign depends on the direction of $\gamma_{\ast j}$.

The saddle points with ${\rm Im}[k]=0$ lead to the terms that have no exponential decay as $x$ grows. 
Conversely, the saddle points with ${\rm Im}[k] \ne 0$ and participating in the expansion (\ref{theorem2v}) 
yield terms having an exponential decay. If $x$ is not very large, these terms, however, cannot be ignored.

\end{enumerate}



\subsection{Definition of the carcass}
\label{subsec51}

In the previous section we discussed an application of the saddle point method for a single real value of~$V$. 
Here we study the whole family of problems indexed by $V$, where $V$ takes all real values. Our aim is a) to trace the motion of the saddle points $p_{\ast j}(V)$ on $\gH$, and b) to describe the rebuilding of the saddle point contours at 
the catastrophical values of~$V$.

For this, we introduce the concept of {\em the carcass of dispersion diagram} and of {\em active points of the carcass}. The carcass of dispersion diagram is a set of all saddle points of $\gH$ for all possible values of $V$, i.~e.~it is a set of points of $\gH$, on which the group velocity is real. 
An active point is a point of the carcass that participate in the representation (\ref{theorem2v}) with the coefficient $a = \pm 1$ for some positive~$V$. Thus, knowing the carcass and its active points makes building the representation (\ref{theorem2v}) easy.

We remind that $v_{\rm gr}$ is called the group velocity, since if it is found on the real dispersion diagram it corresponds to the velocity of a peak of a narrow-band pulse in the waveguide. Hear, however, we use the term ``group velocity" in the formal sense, just for a function defined by (\ref{vgr}).

Since the dispersion equation (\ref{e2204}) imposes two real restrictions, and the equation
\begin{equation}
\label{carcassEq}
\Im[v_{\rm gr}(p)] = 0
\end{equation}
imposes a single real restriction, the carcass is formed by some lines of real dimension 1.

We assume that the points at which $\ptl_k\omega = 0$ and the points at which $\ptl_{\omega}k = 0$ belong to the carcass. They are the points at which the group velocity is equal to zero and to infinity, respectively.

Obviously, the points of $\gH$ with real $\omega$ and $k$ belong to the carcass of the dispersion diagram. Thus, due to the reality statement, all preimages of the real $k$-axis on $\gH$ belong to the carcass. 

We declare that the points of the carcass with real $k$ and $\omega$ typically contain ``not enough information", and the whole complex dispersion diagram, i.e.\ the whole surface $\gH$ contains ``too much of information", while the carcass contains a proper amount of information, just enough to describe pulses in the waveguide.

Let us study the carcass locally. Consider a small neighborhood on $\gH$ with a local complex variable $z$. Let some point $z_0$ belong to the carcass. There are three cases that should be studied:
\begin{itemize}
\item
Case I:
\begin{equation}
\label{vgr1a}
\partial_z v_{\rm gr}(z_0) \neq 0;
\end{equation}
\item
Case II:
\begin{equation}
\label{vgr1b}
\partial_z v_{\rm gr}(z_0) = 0, \qquad
\partial_z^2 v_{\rm gr}(z_0) \neq 0;
\end{equation}
\item
Case III:
\begin{equation}
\label{vgr1c}
\partial_z v_{\rm gr}(z_0) = 0, \qquad
 \partial_z^2 v_{\rm gr}(z_0) = 0.
\end{equation}
\end{itemize}
We consider the cases I and II in details, and assume that Case~III is unstable, i.e.\ it can be destroyed by a small variation of the waveguide parameters. If such variation is made, the point of Case III becomes broken into several closely located points belonging to Case~II.

In Case I, the solution of (\ref{carcassEq}) in the selected neighborhood yields a single line passing through~$z_0$. To illustrate this we substitute the Taylor series
\begin{equation}
\label{Taylor1}
v_{\rm gr}(z) \approx v_{\rm gr}(z_0) + \ptl_z v_{\rm gr}(z_0)\, (z-z_0)
\end{equation}
($z$ belongs to the carcass) into (\ref{carcassEq}). Since $v_{\rm gr}$ is real on the carcass,
\begin{equation}
\Im[\ptl_z v_{\rm gr}(z_0) \, (z - z_0)] = 0.
\end{equation}
The latter is the equation of a line 
\begin{equation}
\label{line_case1}
\alpha \Re[z - z_0] + \beta \Im[z - z_0] = 0,
\end{equation}
with real coefficients $\alpha = \Im[\ptl_z v_{\rm gr}(z_0)]$ and $\beta = \Re[\ptl_z v_{\rm gr}(z_0)]$. The value of $v_{\rm gr}$ grows or decays monotonically on the line (\ref{line_case1}). 

In Case II, the solution of equation (\ref{carcassEq}) yields two lines passing through $z_0$. This also can be illustrated by the Taylor series
\begin{equation}
\label{Taylor2}
v_{\rm gr}(z) \approx v_{\rm gr}(z_0) + \frac{1}{2} \ptl^2_z v_{\rm gr}(z_0)\, (z-z_0)^2.
\end{equation}
One can say that four half-lines of the carcass are attached to each other at such a point.

\begin{figure}
\centering
\includegraphics[scale = 1]{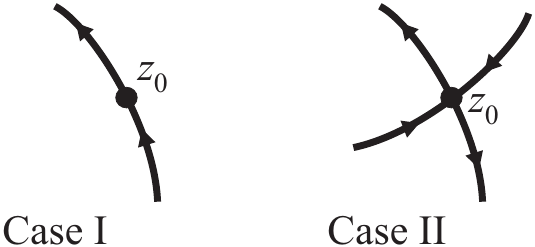}
\caption{Fragment of the carcass in Case I and Case II}
\label{figB3102}
\end{figure}

Sketches of fragments of the carcass in Case I and Case II are shown in \figurename~\ref{figB3102}. The arrows show the direction of growth of $v_{\rm gr}$.

If the local variable is $k$, then the conditions of Case II are as follows:
\begin{equation}
\label{condCase2}
\frac{d^2 \omega}{dk^2} = 0, \qquad \frac{d^3 \omega}{dk^3} \neq 0.
\end{equation}
If $(\omega, k)$ belong to the real dispersion diagram, then conditions (\ref{condCase2}) correspond to the {\em inflection point} of the dispersion diagram in a clear geometrical sense.

Below we call {\em any} point $p \in \gH$ with
\begin{equation}
\label{eq425}
\ptl_z v_{\rm gr}(p) = 0, \quad \ptl_z^2 v_{\rm gr}(p) \neq 0
\end{equation}
an inflection point ($z$ is a local variable). The inflection points on which the group velocity is real are crossings of the lines of the carcass (\figurename~\ref{figB3102}, Case II). 


Note that
\begin{equation}
\frac{d^2k}{d\omega^2} = \frac{d}{d\omega} \frac{1}{d\omega/dk} = - \left( \frac{d\omega}{dk} \right) ^{-2} \frac{d^2 \omega}{dk^2}.
\end{equation}
Thus, if $\ptl _{\omega}k \neq 0$ and $\ptl_{k} \omega\neq 0$, conditions
$$\frac{d^2 \omega}{dk^2}=0$$
and
$$\frac{d^2 k}{d\omega^2}=0$$
are equivalent. Any one of them can be used for finding the inflection points.

Define also the inflection points at infinity. One can define a compactification of $\gH$ by adding the infinitely remote points and studying $v_{\rm gr}$ at those points. A local variable for an infinite point is
\begin{equation}
z =\frac{1}{k}.
\end{equation}
Then one can check whether the condition (\ref{eq425}) is fulfilled or not. It follows from the asymptotic study of (\ref{e2204}) that the dispersion diagram has the following asymptotic near the infinity points: 
\begin{equation}
\label{eq426}
\omega(k) = \pm c_ik+\alpha_i k^{-1} + O(k^{-3})
\end{equation}
for some real $c_i$,  $\alpha_i$, and $k$. Indeed $D(\omega,k)$ depends only on even orders of $\omega$ and $k$. This fact is obvious when $\gD_1=0$, and follows from (\ref{e2103}) when $\gD_1$ is not zero.

 Then,
\begin{equation}
\label{eq427}
v_{\rm gr}(k) = \pm c_i-\alpha_i k^{-2} + O(k^{-4}),
\end{equation}
thus two lines of carcass (the remote parts of the real and the imaginary axis of $k$) cross at
infinity, so it should behave as an inflection point there.


%
%

Let us summarize the facts about the carcass that follow from the discussion above.
\begin{itemize}
\item
The saddle points of representation (\ref{theorem2v}) belong to the carcass. However, not each point of the carcass corresponding to some value $V$ participates in this representation with $a_j \neq 0$.
\item
The carcass is composed of several lines on $\gH$.
\item
The preimages of the real $k$-axis on $\gH$ belong to the carcass. Corresponding lines of the carcass are the {\em real branches} of the carcass. There are, possibly, some other branches of the carcass (the {\em complex branches}) that also play some role in wave processes. Points of the carcass on its complex/real branches correspond to wave components with/without an exponential decay.
\item
At each inflection point (of Case II, (\ref{vgr1b}), \figurename~\ref{figB3102}) with real $v_{\rm gr}$ there are four branches 
of the carcass that meet each other. For example, some complex branches of the carcass are attached to real branches at the  geometrical inflection points of the latter. The inflection points on the real dispersion diagram are local maximums and minimums of the group velocity.
\item
The infinities on $\gH$ are inflection points in a certain sense, i.e.\ four branches of the carcass meet each other at each such point. 
\item
The group velocity $v_{\rm gr}$ is a monotonic function on the segments of lines between the ``bad'' points of the carcass that are inflection points and the points with $v_{\rm gr} \to \pm \infty$. This is a trivial consequence of the fact that everywhere on the carcass outside the ``bad'' points $\ptl_z v_{\rm gr}$ has a non-zero finite value.
\item
Similarly to $v_{\rm gr}$, the inverse group velocity $v_{\rm gr}^{-1}$ is a monotonic smooth function along the carcass outside the points, where $\ptl_z v_{\rm gr}^{-1} = 0$ or $v_{\rm gr}^{-1} \to \infty$. This property helps us to build  the carcass near the points with $v_{\rm gr}\to \infty$.  



\end{itemize}
For practical calculations, ``knowing the carcass'' on $\gH$ is possessing a dense enough array 
of quadruplets $(\omega , k, v_{\rm gr} , {\rm f})$, where $(\omega , k)$ are the coordinates of the point of the carcass, 
$v_{\rm gr}$ is the group velocity at the point, and ${\rm f}$ is a binary flag, equal to ``true'' if the point is active and to ``false''
if not.

Building of the carcass is not a complicated task from the numerical point of view. 
Namely, if some point  $(\omega , k, v_{\rm gr} , {\rm f})$ of the carcass is known, one can find a point 
located near it and having the group velocity $v_{\rm gr} + \delta v$ for some small real $\delta v$. 
For this, one should solve the system 
\[
D(\omega' , k' ) = 0 , 
\qquad 
\frac{\ptl_k D (\omega' , k' )}{\ptl_\omega D (\omega' , k' )} = - (v_{\rm gr} + \delta v),
\]
say, by Newton's method, taking $(\omega , k)$ as the starting values. The flag ${\rm f}'$
of the point $(\omega' , k')$ is then found by the 
classification algorithm described in the next subsection. 

Thus, to compute the carcass, one should find at least one point on each branch of it. 
This may be a non-trivial task, but is simplified by the fact that most branches are connected at the 
inflection points. So one can search for complex branches of the carcass in neighborhoods of the inflection points.

\subsection{Finding active points on the carcass}
\label{subsec52}

Return to the representation (\ref{e3106}). Fix a positive real value $V$ and apply the procedure of contour deformation described in the proof of the theorem. As the result, get the set of saddle points $p_{\ast j}$ and saddle point contours $\gamma_{\ast j}$ from (\ref{theorem2v}). 
As we already mentioned, the points $p_{\ast j} (V)$, $V > 0$ 
for which $a_j \neq 0$ are referred to as {\em active}. Here we describe 
the process of classification of the points of carcass, i.e.\ determining whether a point is active or not. 


Some points of the carcass can be classified in an elementary way. 
All points with real $k$ and positive value of $v_{\rm gr}$ are active.
This follows from the procedure of building of the steepest descend contours.   
The points with $v_{\rm gr} < 0$ are  not active by definition. 
Let $c$ be the maximum eigenvalue defined from the problem (\ref{e2402}). 
Then all points with 
$v_{\rm gr} > c$ are not active.   
This follows from the fact that the field is identically equal to zero for $x/ t > c$. 
If $g_i(p{_\ast}) < 0$, where $p_{\ast}$ is a point of the carcass, then the point $p_{\ast}$ is not active.  

One should classify the points of the carcass not covered by the cases listed above.
The procedure of deformation of the initial integration contours described in the proof of the theorem enables one to formulate the following algorithm of establishing the activity of some point $p_{\ast}$ belonging to the carcass. 
For this, we are trying to reverse the procedure, i.e.\ to go along the streams in the opposite directions, and to check, whether a preimage of the real $k$-axis (the initial integration contour) is hit.

\vskip 6pt
\noindent
{\bf Algorithm of classification of the carcass points}

For a given value of $V$ define the function $g_i(p;V)$ by (\ref{e4102}) on $\gH$. Let $p_{\ast}$ be a saddle point for $g_i$. 
Find the value $g_i(p_{\ast};V)$. 
Let be $g_i(p_{\ast};V) > 0$. 
Draw two streams that go {\em to\/} 
the point $p_{\ast}$, and follow these streams from $p_{\ast}$ in the direction opposite to their orientation, i.e.\ go along the streams in the direction of decay of~$g_i$. Find the points $p_1$ and $p_2$ of these streams, for which $g_i(p_{1,2};V) = 0$. Let be $p_{1,2} = (\omega_{1,2}, k_{1,2})$. Three cases can happen:
\begin{enumerate}
\item
$\Im[k_1] \neq 0$ and $\Im[k_2] \neq 0$ (none of the streams hits the real $k$-axis);
\item
$\Im[k_1] = 0$ and $\Im[k_2] = 0$ (both streams hit the real $k$-axis);
\item
$\Im[k_1] = 0$ or $\Im[k_2] = 0$  (just one stream hits the real $k$-axis, but not both of them).
\end{enumerate}
The point $p_{\ast}$ is active in case~3 
and not active in cases 1 and~2.

If $p_{\ast}$ is active, take the streams going from the point $p_{\ast}$ and compose the saddle point contour $\gamma_{\ast}$ of them. Assign the orientation of $\gamma_{\ast}$ according to \figurename~\ref{fig4_4}. With this orientation, the coefficient $a$ corresponding to this contour in (\ref{theorem2v}) will be equal to 1.

\section{Numerical demonstrations}
\label{sec6}

\subsection{Carcass for the motivating example}
\label{subsec61}

Let us build the carcasses and their reduced counterparts for some waveguides. First, let us study the carcass for the motivating example of the section~\ref{subsec24}. The dispersion equation for this system is a polynomial relation of the fourth order with respect to $\omega$ and $k$. Thus, it can be represented as a function $\omega(k)$ on a 4-sheeted surface. 

\begin{figure}
\centering
\includegraphics[scale = 0.8]{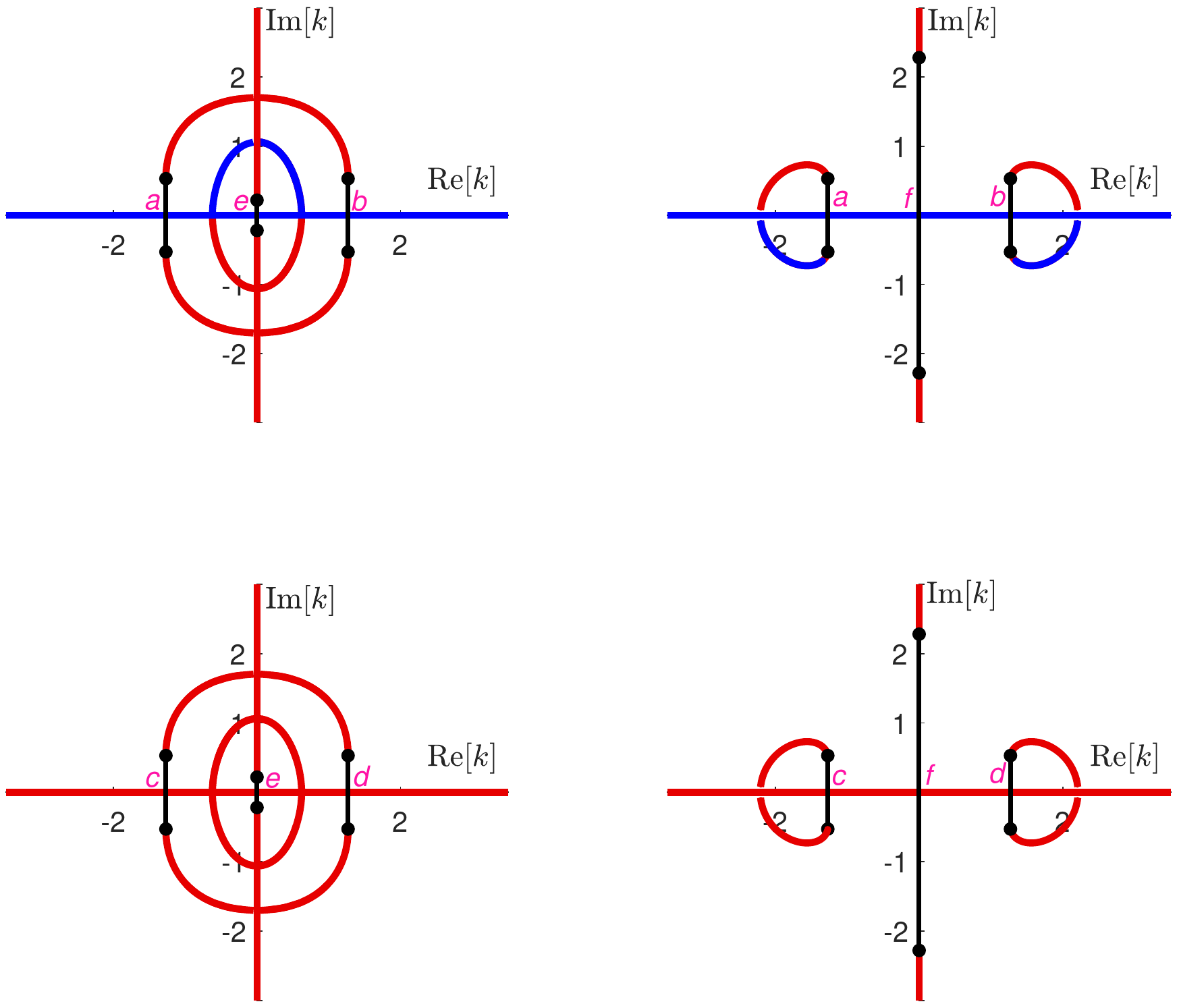}
\caption{Carcass for the motivating example. Blue: active points, red: not active points. The cuts of $\gR$ sheets, which are attached to each other are denoted by  pink letters}
\label{Lcarcass}
\end{figure}

The Riemann surface $\gR$  for the system 
and the carcass on it are shown  in Figure~\ref{Lcarcass}. The bold black points are branch points and the black 
lines are branch cuts. The sides of the cuts that are attached to each other are denoted by the same
pink letters. The red and blue lines on $\gR$ form the carcass. Active points of the carcass are blue, and the
not active points are red. 


The upper sheets in the figure correspond to  
the modes travelling in the positive direction. 
The branches  of the carcass with real $k$  
provide the stationary phase asymptotics. 

One can see that there are some branches of the carcass with active points 
having complex~$k$. According to the consideration above, these branches provide 
saddle point terms that are not described by the stationary phase method
and that correspond to exponentially decaying waves. In particular, the blue arc 
in the top-left part of \figurename~\ref{Lcarcass} results in the correction term shown 
in \figurename~\ref{fig03},~left. 
  
The complex saddle points emerge near the inflection points of the dispersion 
diagram (the maxima and minima of the group velocity). 
The saddle point method fails near the inflection points, and the field should be described by an Airy-type 
asymptotics~\cite{Pekeris1948}.
For example, the saddle point asymptotic terms should be replaced by the Airy asymptotics for $17< t <19$ 
in \figurename~\ref{fig03},~left.  


The lower sheets in the figure describe  waves going in the negative direction. All points 
of the carcass on 
these sheets are not active, since $v_{\rm gr}<0$ there.




\subsection{Other types of carcasses for 2D systems}
\label{AppB}

Not all carcasses of 2D systems are similar to the one that is shown in the Figure~\ref{Lcarcass}.  
Below we provide some examples.


Consider the  WaveFEM equation (\ref{e2101}) with:
\begin{equation}
\label{e6201}
\begin{gathered}
\gM = {\rm I}, \quad
\gD_2 = \begin{pmatrix} 4 & 0 \\ 0 & 1 \end{pmatrix}, \quad
\gD_1 = 0, \quad
\gD_0 = \begin{pmatrix} -15 & -2 \\ -2 & -45 \end{pmatrix}
\end{gathered},\quad \gF = \left( \begin{array}{c}
1 \\
0
\end{array} \right).
\end{equation}

\begin{figure}[h!]
\centering
\includegraphics[width = 8cm]{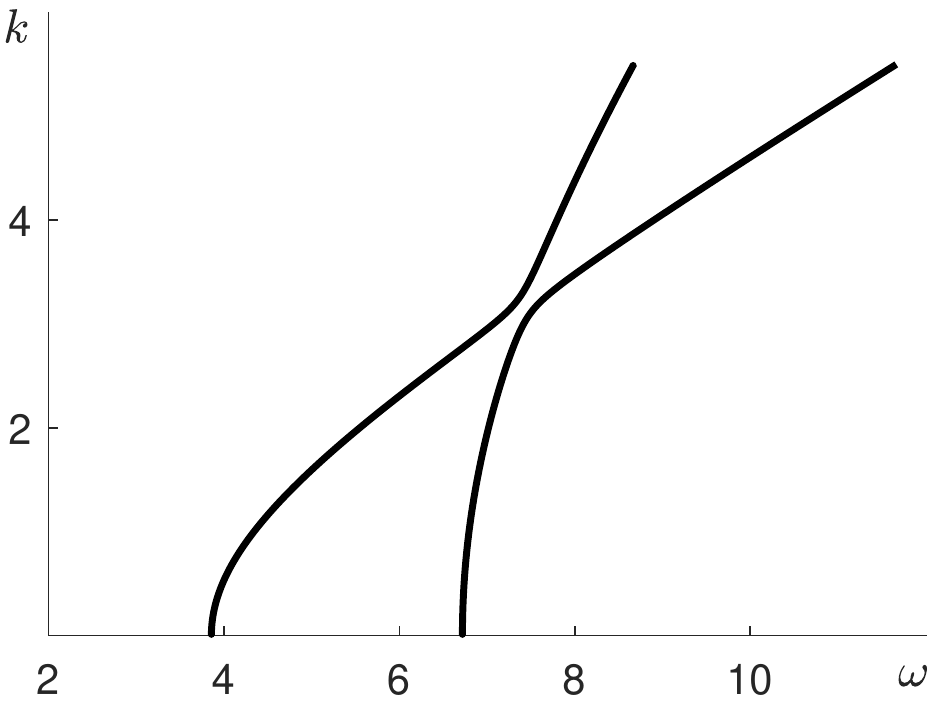}
\hspace{4ex}
\includegraphics[width = 8cm]{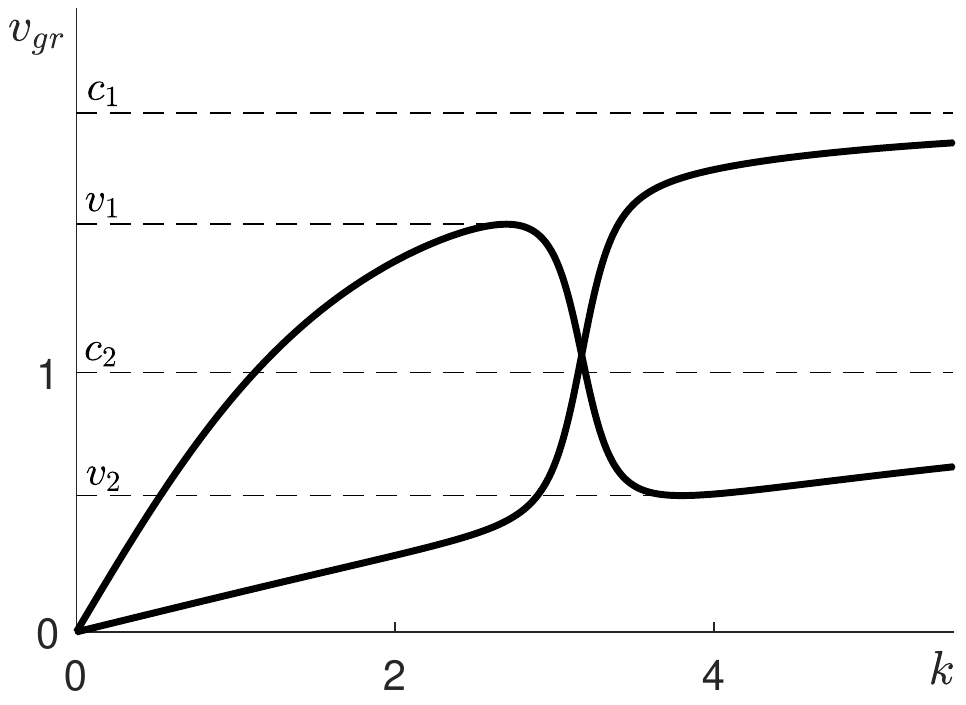}
\caption{WaveFEM equations with (\ref{e6201}):  dispersion diagram (left) and group velocities (right)}
\label{DDB}
\end{figure}

Only $\gD_0$ has non-diagonal terms, and they are small comparatively to the diagonal terms. 
Thus, the WaveFEM equation describes a physical system of two scalar waveguides weakly coupled with each other.  
The real dispersion diagram and dependence of the group velocity on $k$ are presented in \figurename~\ref{DDB}. 
One can see that the real dispersion diagram displays a well-known behavior of avoiding crossing of the branches. 
This is a typical behavior of a system with a weak coupling.   


\begin{figure}
\centering
\includegraphics[scale = 0.8]{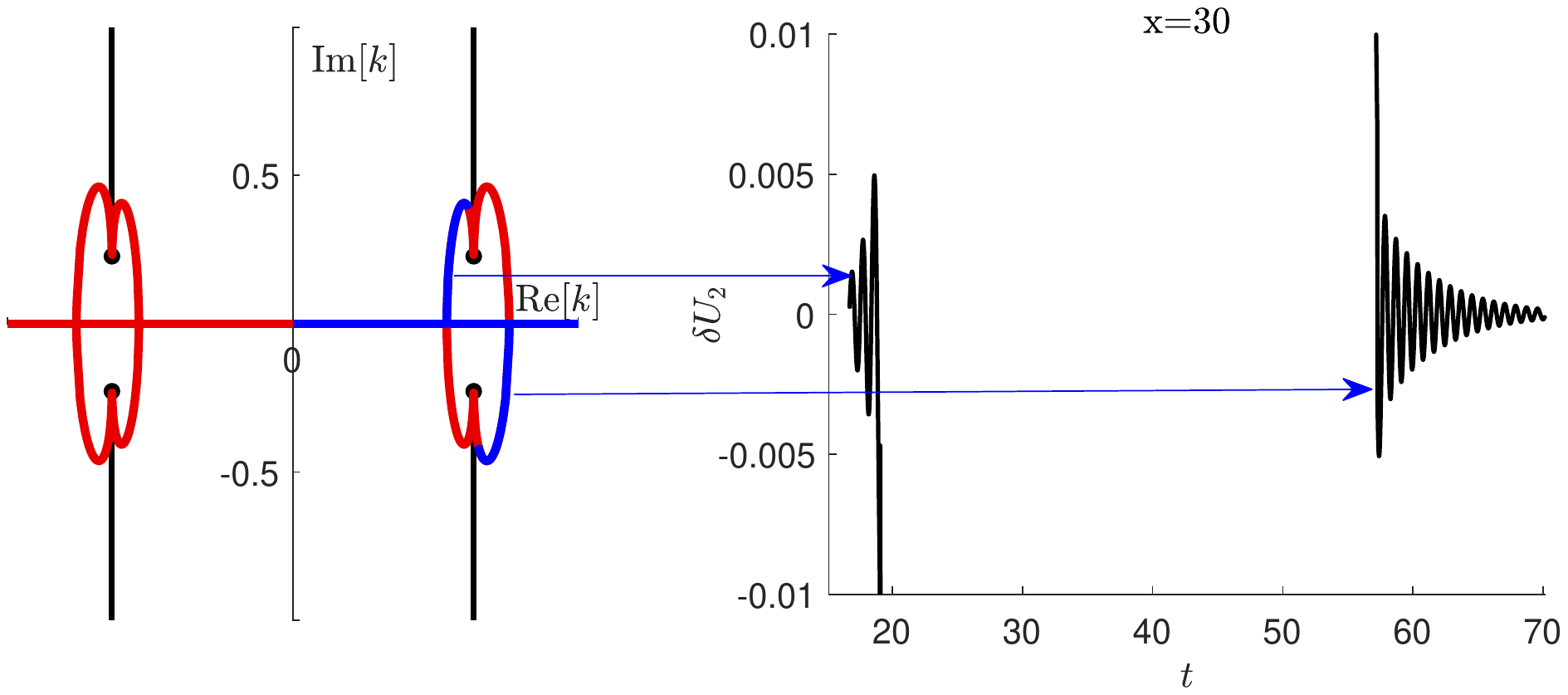}
\caption{Part of the carcass corresponding to the second mode of the system (\ref{e6201}) (left) and the field formed by active points with $\Im[k] \neq 0$ (right). Blue/red -- active/not active points of the carcass} 
\label{carcass_B}
\end{figure}

A carcass (to be more precise, an important part of the carcass belonging to one of the sheets of $\gR$) for this waveguide is presented in \figurename~\ref{carcass_B},~left. 
One can see that  the surface $\gR$ has branch points located near the wavenumber of the avoiding crossing.
A part of the carcass (a complex branch) passes through each branch point. Some portions of such branches
are active (blue), thus they correspond to exponentially decaying wave pulses.   
These pulses are shown in the right part of the figure. 
Arrows indicate the links between the parts of the carcass and the pulses. 
Note that we demonstrate only the additional pulses; the components provided by the stationary phase method 
are not shown.

Note also that the topological structure of the carcass for the matrices (\ref{e6202}) is different from that of (\ref{e2401}). 
Besides, the additional pulses are different. For example, just the active parts of complex branches of the carcass 
for (\ref{e6202}) are relatively short and non closed, so one can expect narrow band pulses. Some further discussion 
can be found in \cite{ShaninToyama}
where it is shown that such a coupling between subsystems leads  
to what is called the {\em exchange pulse}. 

\begin{figure}
\centering
\includegraphics[width = 8cm]{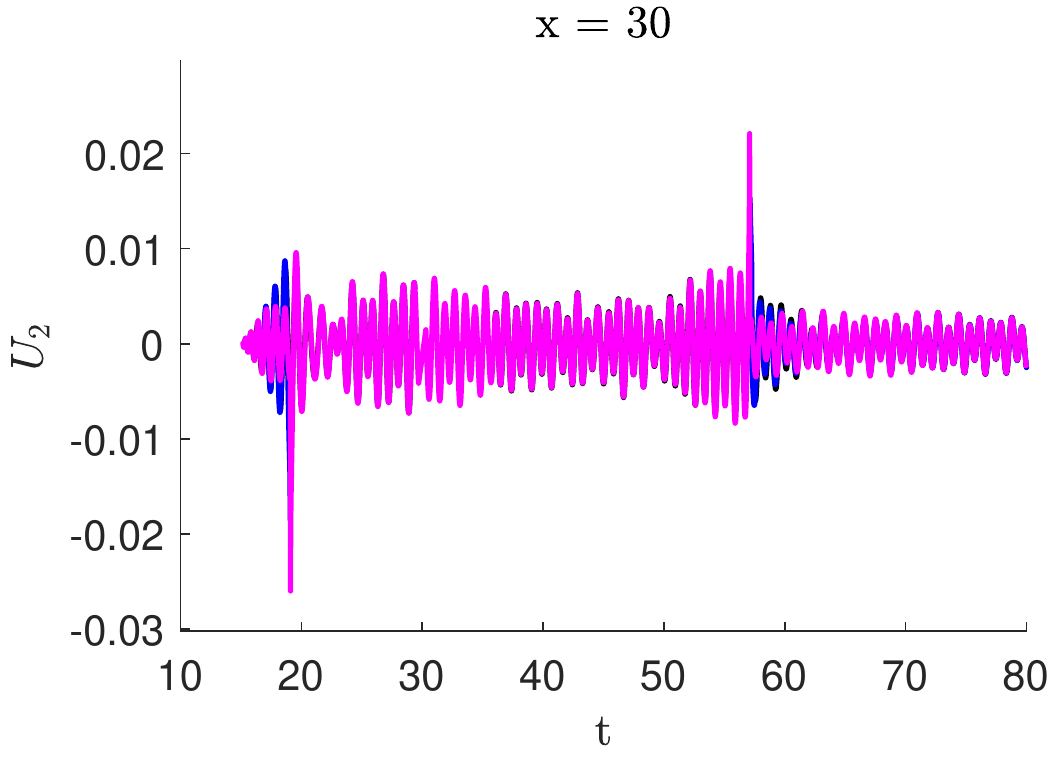}
\\
\includegraphics[width = 8cm]{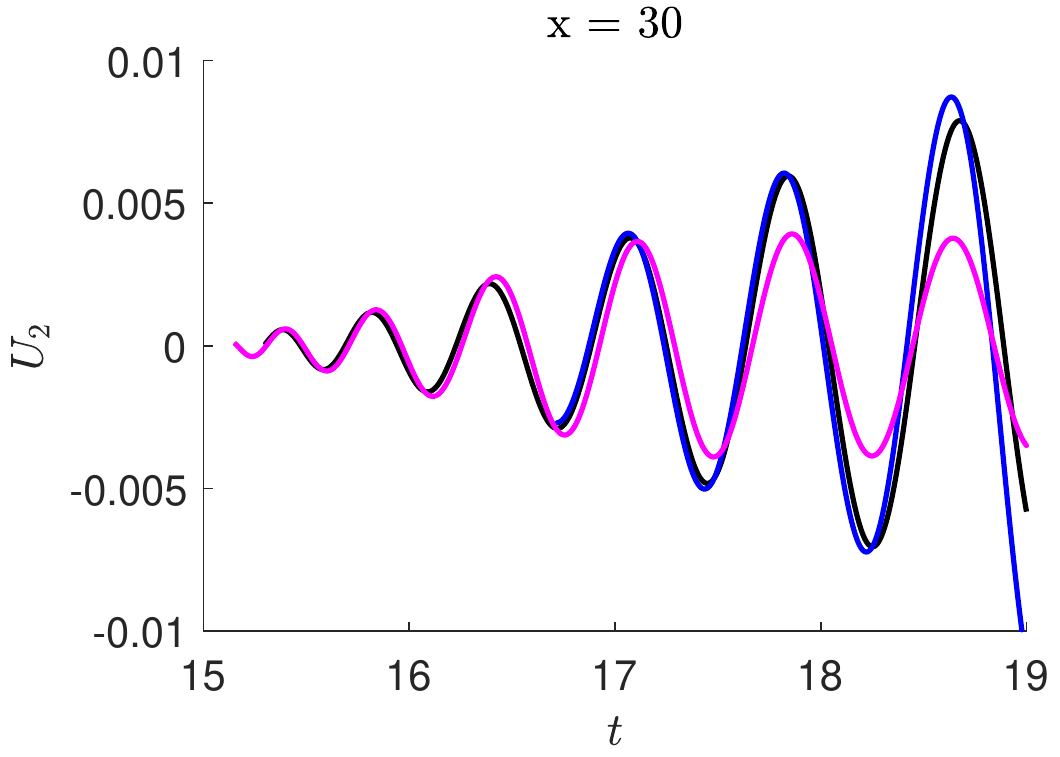}
\includegraphics[width = 8cm]{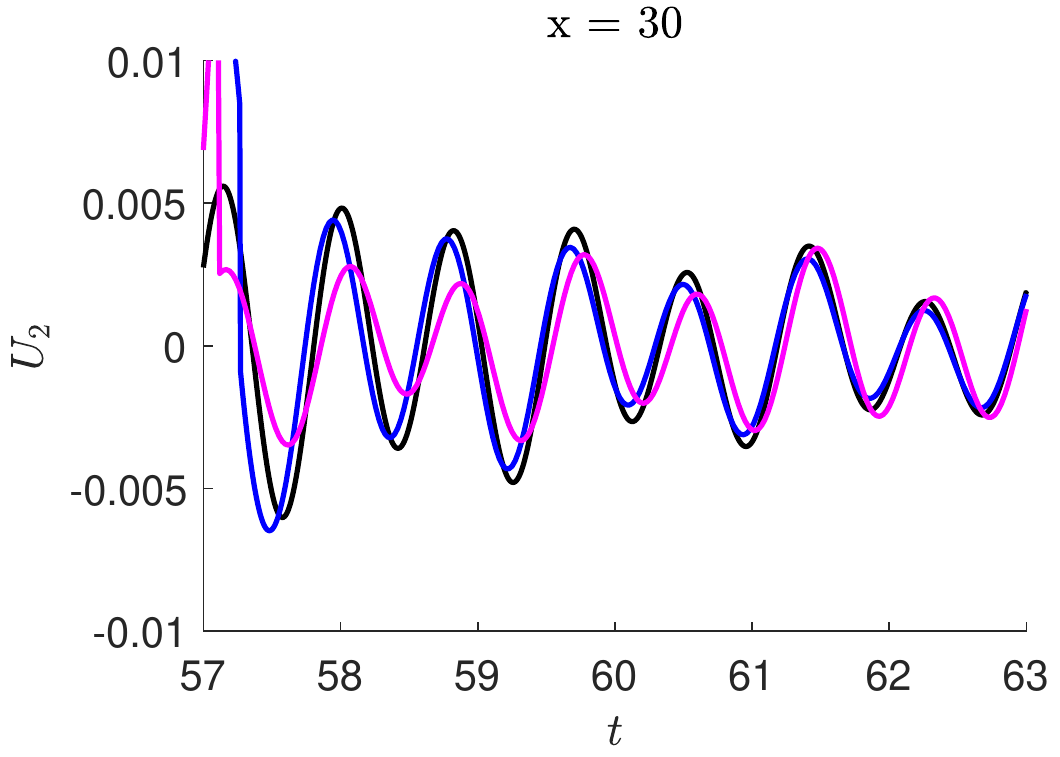}
\caption{Pulse response of the system (\ref{e6201}). Black: numerically calculated integral (\ref{e2205}), magenta: stationary phase method, blue: saddle point method. The bottom part of the figure shows the zoomed version of the top part}
\label{field_B}
\end{figure}

Let us compare the integral (\ref{e2205}) with its stationary phase  and saddle point  asymptotics for the considered system (\ref{e6201}).  The dependence of the field component $U_2$  on time for a fixed coordinate ($x = 30$) is presented in \figurename~\ref{field_B}. The result of numerical estimation of integral (\ref{e2205}) is shown in black,  the stationary phase asymptotic is shown in magenta, and  the saddle point approximation is shown in blue. Sharp peaks of the black line near $t = 30$ 
are connected with the inaccuracy of the numerical calculations near one of the limiting velocities $c_2 = 1$, and sharp peaks of  magenta and blue lines correspond to local extremums of the group velocities, i.e.\ to the case, when the saddle point asymptotic is not valid. If we do not consider these segments, all three lines in \figurename~\ref{field_B} coincide except for the areas near $t<20$ and $55<t<60$, i.e.\ the areas, where  pulses  provided by complex saddle points  (see~\figurename~\ref{carcass_B}) have significant amplitude. These areas are zoomed  in the bottom part of the figure.


A more exotic example is given by the WaveFEM equations with
\begin{equation}
\label{e6202}
\begin{gathered}
\gM = {\rm I}, \quad
\gD_2 = \begin{pmatrix} 4 & 0 \\ 0 & 1 \end{pmatrix}, \quad
\gD_1 = 0, \quad
\gD_0 = \begin{pmatrix} -24 & -2 \\ -2 & -24 \end{pmatrix},\quad \gF = \left( \begin{array}{c}
1 \\
0
\end{array} \right).
\end{gathered}
\end{equation}

\begin{figure}[h!]
\centering
\includegraphics[width = 7cm]{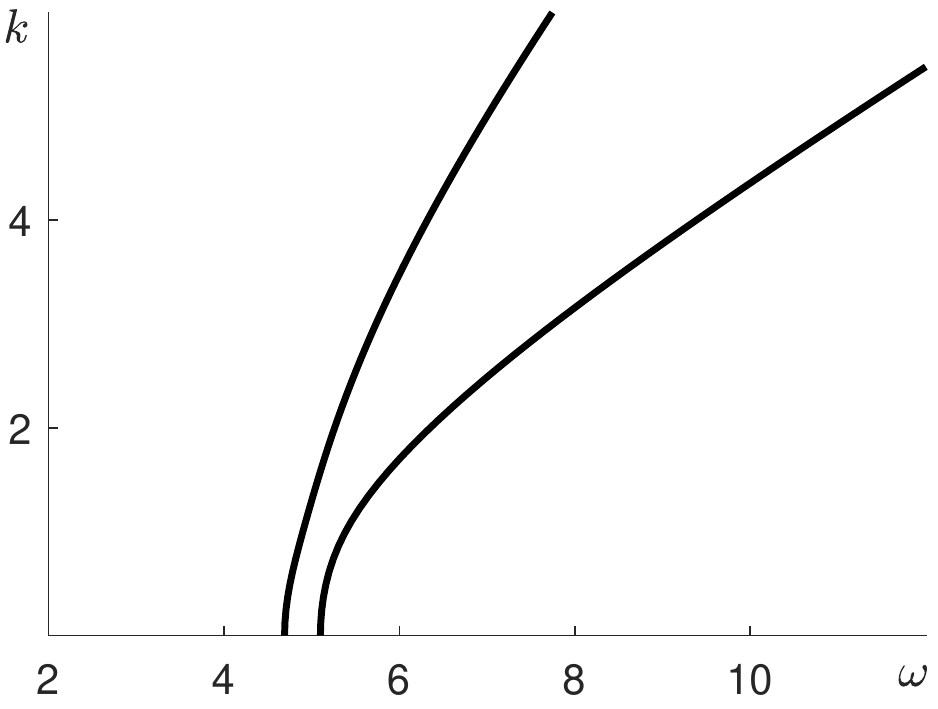}
\hspace{4ex}
\includegraphics[width = 7cm]{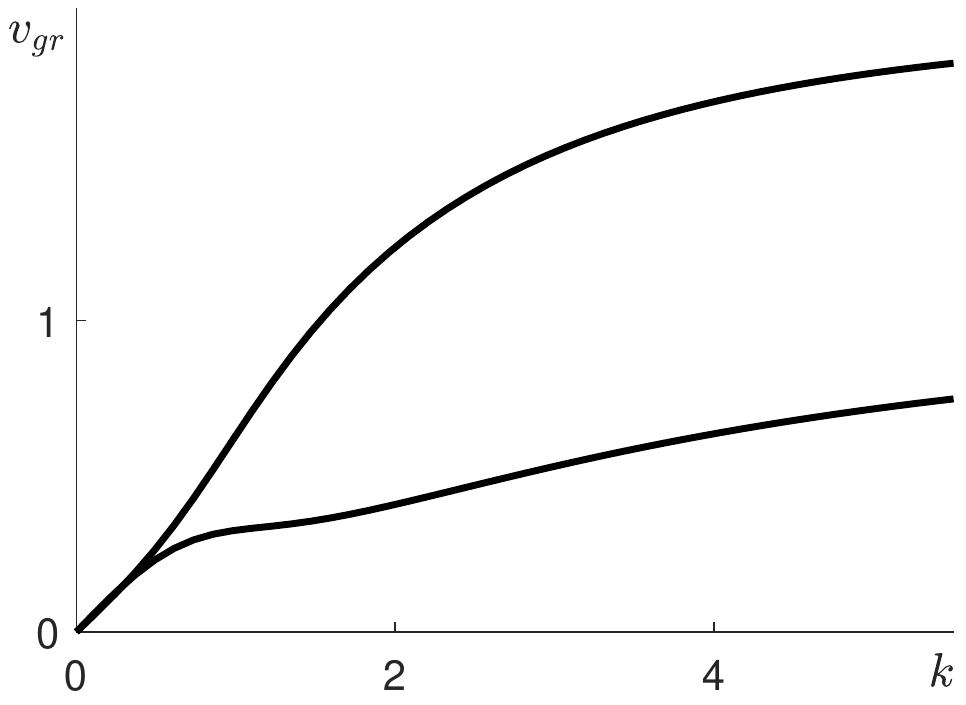}
\caption{Waveguide (\ref{e6202}): real dispersion diagram (left) and  group velocities (right)}
\label{DDA}
\end{figure}

The real dispersion diagram and group velocities are shown in \figurename~\ref{DDA}. The main feature of this system is that there are no local extrema of $v_{\rm gr}(k)$.

\begin{figure}
\centering
\includegraphics[width = 15cm]{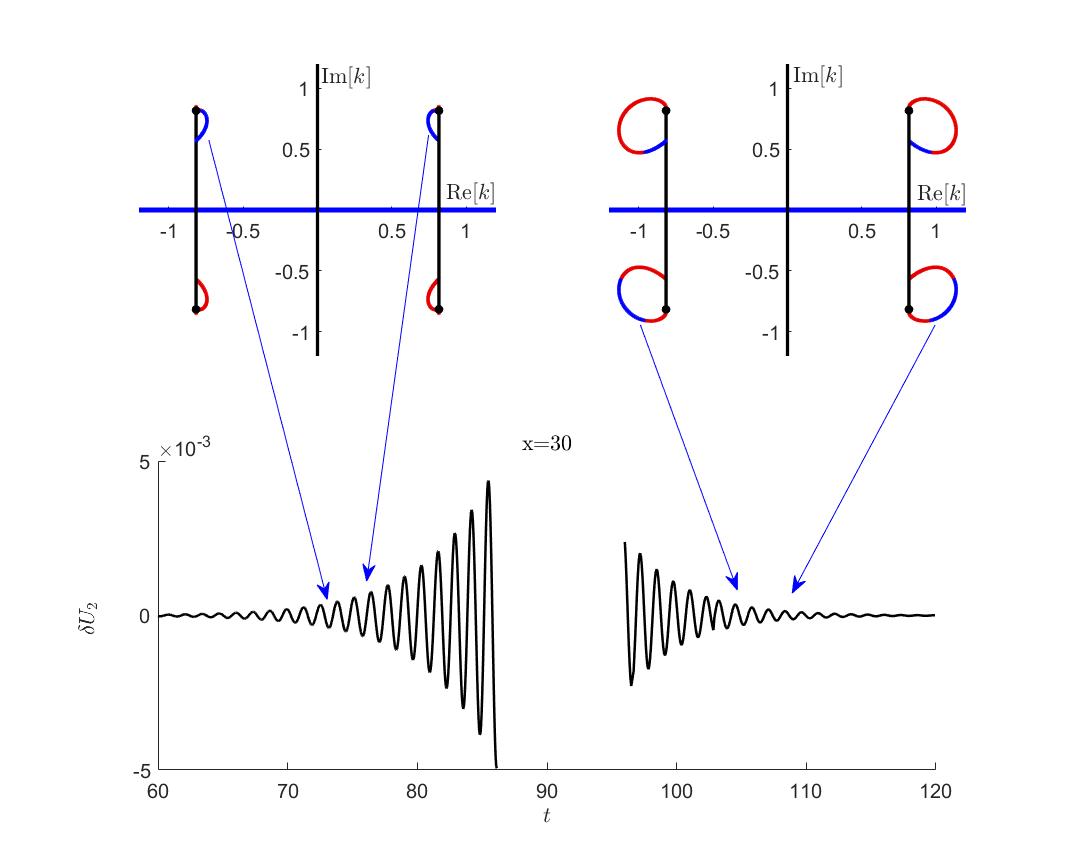}
\caption{Part of the carcass of the system (\ref{e6202}) corresponding to the modes propagating in the positive direction (top) and the field formed by active points with $\Im[k] \neq 0$ (bottom). Blue/red -- active/not active points of the carcass} 
\label{carcass_A}
\end{figure}

An interesting  part of the carcass is shown in \figurename~\ref{carcass_A}, left.
The complex branches of the carcass do not intersect with the real branches. However, there are some active parts on 
the complex branches, so one should consider the saddle points from them when constructing the saddle point asymptotics.  
These points form decaying pulses, which are shown in the bottom part of \figurename~\ref{carcass_A}.
 

\begin{figure}[h!]
\centering
\includegraphics[width = 8cm]{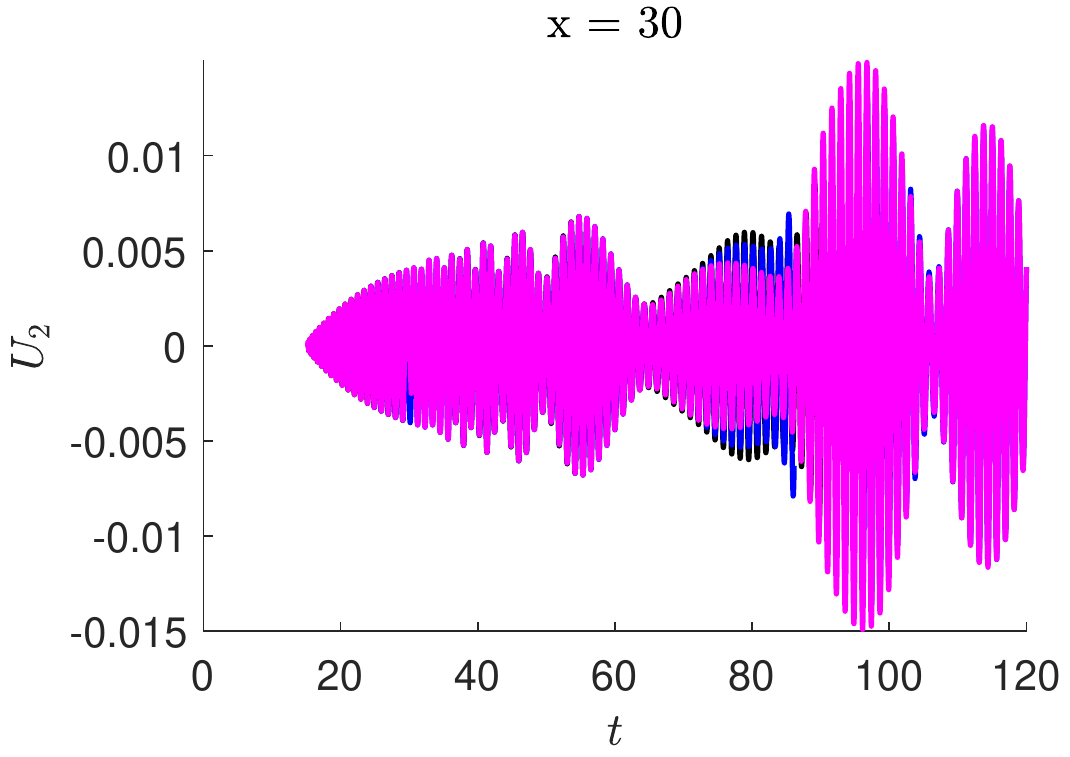}
\hspace{4ex}
\includegraphics[width = 8cm]{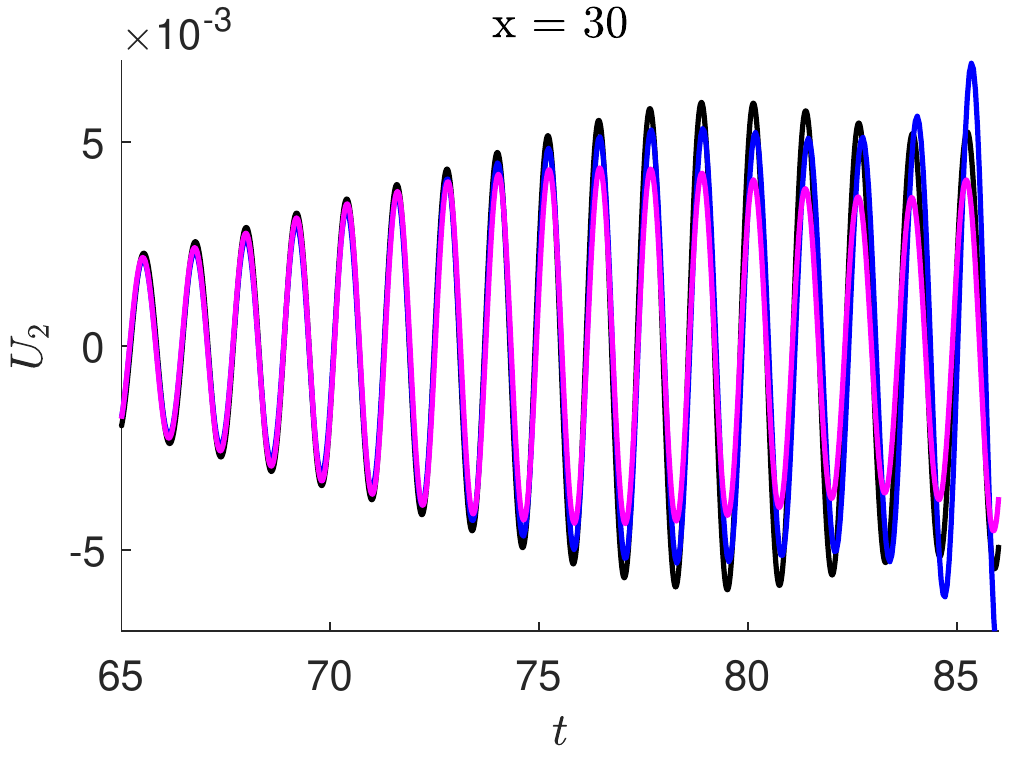}
\caption{Pulse response of the WaveFEM with (\ref{e6202}): (left) large time interval, (right) a zoomed fragment. Black: numerically calculated integral (\ref{e2205}), magenta: stationary phase method, blue: saddle point method}
\label{fieldA}
\end{figure}

In \figurename~\ref{fieldA} we show a comparison of the numerically calculated wavefield and the results of the application of the stationary phase method and the saddle point method. Again, the saddle point method allows one to obtain a 
better estimation  of the wavefield for such values of $t$, where the amplitude of the additional pulses (\figurename~\ref{carcass_A}, right) is comparable to the amplitude of the wavefield.

\subsection{A carcass for WaveFEM equations of a higher dimension. A forerunner}
\label{subsec63}

\begin{figure}
\centering
\includegraphics[width = 8cm]{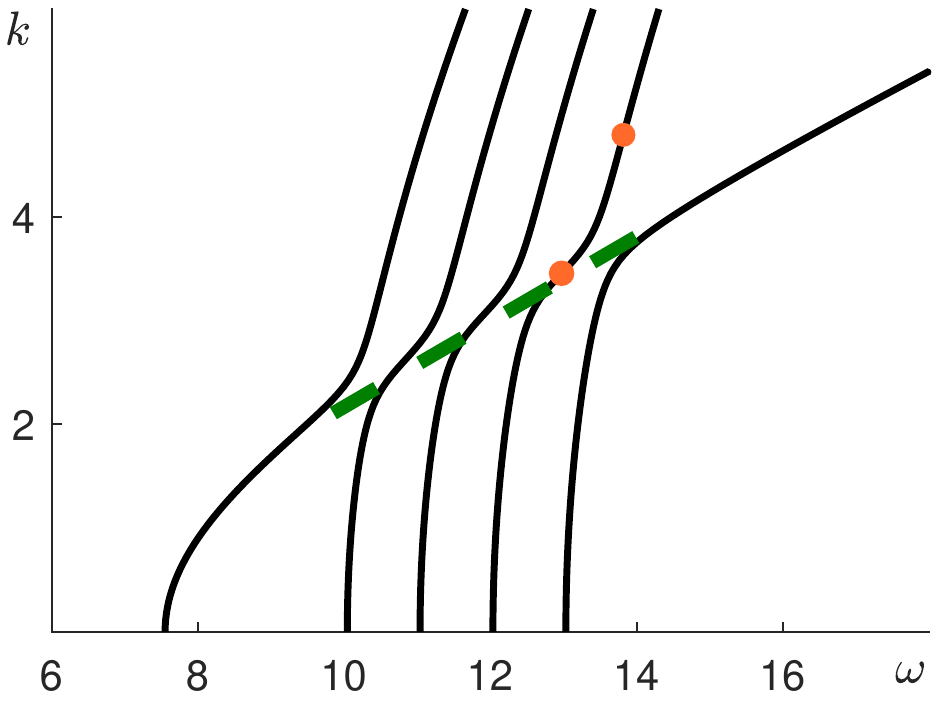}
\hspace{4ex}
\includegraphics[width=8cm]{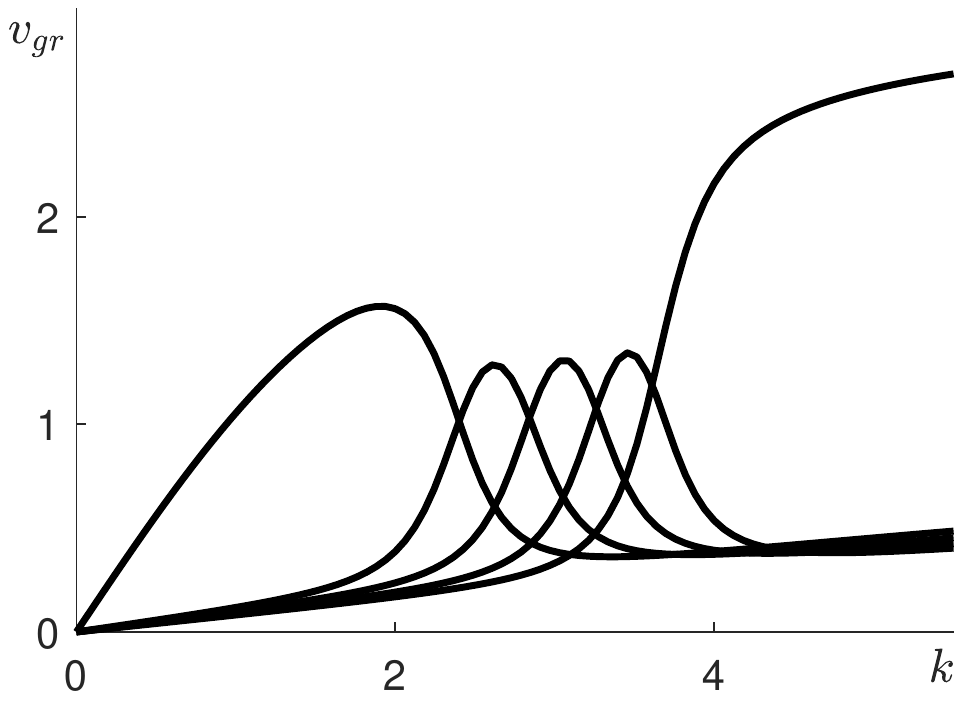}
\caption{Waveguide (\ref{eqB310041}) -- (\ref{eqB310044}): real dispersion diagram (left) and group velocities (right)}
\label{figB31021}
\end{figure}

Consider a layered waveguide comprised of five  weakly coupled scalar subsystems. 
Let the first subsystem bear a fast mode, and the remaining four subsystems bear  slow modes. 
The WaveFEM equation is as follows:
\begin{equation}
\gM = m \,  {\rm I}, \qquad m = 1,
\label{eqB310041}
\end{equation}
\begin{equation}
\gD_2 = \left( \begin{array}{ccccc}
c_2^2 & & & &  \\
 & c_1^2 & & &  \\
 & & c_1^2 & &  \\
 & & & c_1^2 &  \\
 & & & &  c_1^2
\end{array} \right),
\qquad
c_1 = 1, \qquad c_2 = 3,
\label{eqB310042}
\end{equation}
\[
\gD_0 = \gD_0' + \gD_0'',
\]
\begin{equation}
\gD_0' = \left( \begin{array}{ccccc}
-a_1^2 & & & &  \\
 & -a_2^2 & & &  \\
 & & -a_3^2 & & \\
 & & & -a_4^2 &  \\
 & & & &  -a_5^2
\end{array} \right),
\qquad
a_1^2 = 60, \qquad a_m = m+8 \mbox{ for }m>1,
\label{eqB310043}
\end{equation}
\begin{equation}
\gD_0'' = \left( \begin{array}{ccccc}
0 & b_2& b_3 & b_4 & b_5  \\
b_2 & 0& & &  \\
b_3 & & 0& & \\
b_4 & & & 0 &  \\
b_5 & & & &  0
\end{array} \right),
\qquad
b_m = \mu \, (m+4), \qquad
\mu = 1.
\label{eqB310044}
\end{equation}
The term $\gD_1$ is equal to zero. Note that matrices $\gM$, $\gD_0'$, $\gD_2$ are diagonal.  Matrix $\gD_0''$ describes the interaction between the subsystems. Value of $\mu$ is small with respect to $a^2_m$, and thus the interaction is small. The numerical values in the model are taken to provide indicative graphs. The real dispersion diagram is shown in \figurename~\ref{figB31021}.  One can clearly see the dispersion diagram is composed of slow branches, which ``avoid'' crossings with a fast mode (having small slope). The interaction between the fast mode and the family of  slow modes manifests itself in forming the terraced structure in the crossing domain.  The dashed line the figure shows the position of the fast mode in the case of zero interaction between the subsystems. There are many examples of waveguides where such terraced structures can be observed. They are studied in \cite{Mindlin2006} and \cite{Tolstoy1956}, for instance. 

\begin{figure}
\centering
\includegraphics[width = 17 cm]{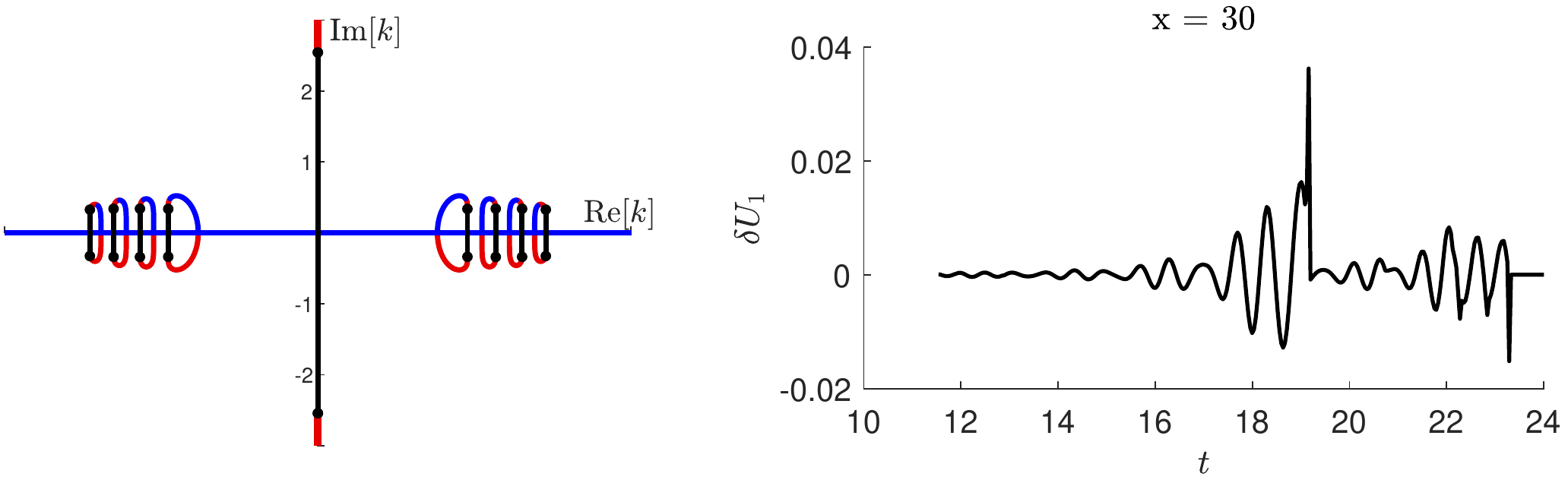}
\caption{Left: part of the carcass of the system (\ref{eqB310041}) -- (\ref{eqB310044}) corresponding to the fast mode: active (blue) and not-active (red) points of the carcass. Black bold lines are cuts, black points are branch points. Right: Pulse that is formed by the complex saddle points of the carcass} 
\label{carcass5D}
\end{figure}

One can see that each branch of the real dispersion diagram except the last one has two inflection points. Such inflection points on the fourth branch are shown in
the figure by orange circles. Since there are 8 inflection points for real positive values of $k$ (four of them correspond to maxima of the group velocity and four points correspond to the minima of the group velocity), one can expect that carcass has 8 branches belonging to the complex domain. The part of the carcass corresponding to the fast mode is shown in \figurename~\ref{carcass5D}. 
The active points in the upper half plane of the 5D system form a pulse, which is shown in \figurename~\ref{carcass5D}, right.

\begin{figure}
\centering
\includegraphics[width = 8cm]{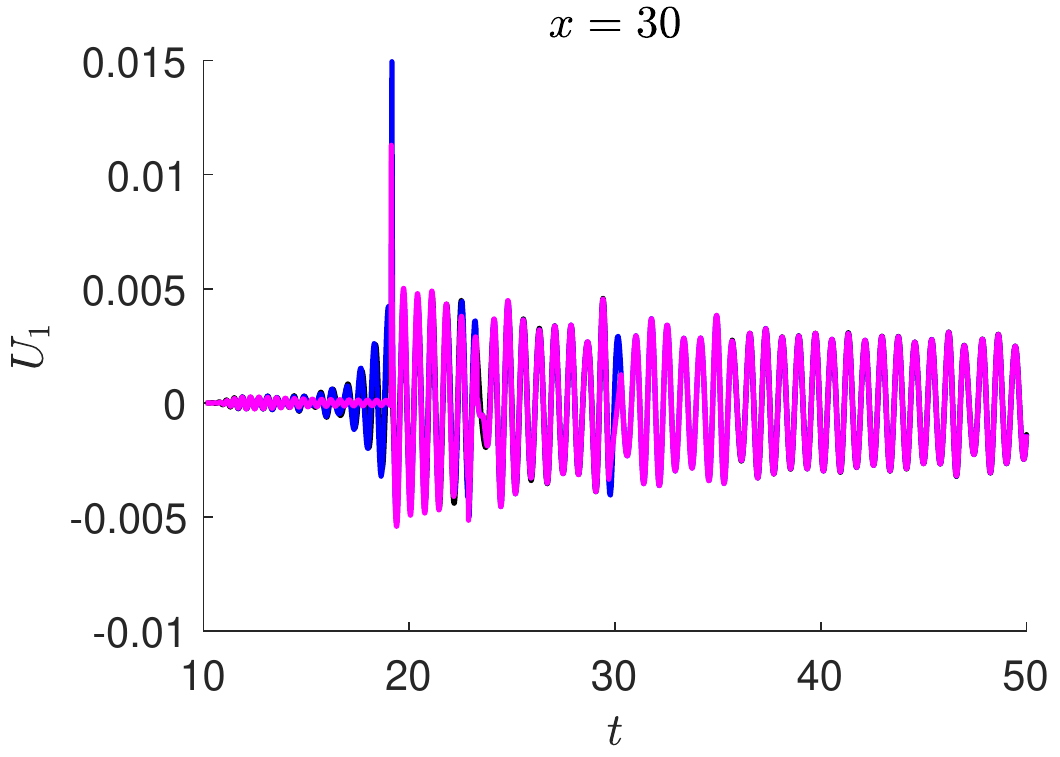}
\hspace{4ex}
\includegraphics[width = 8cm]{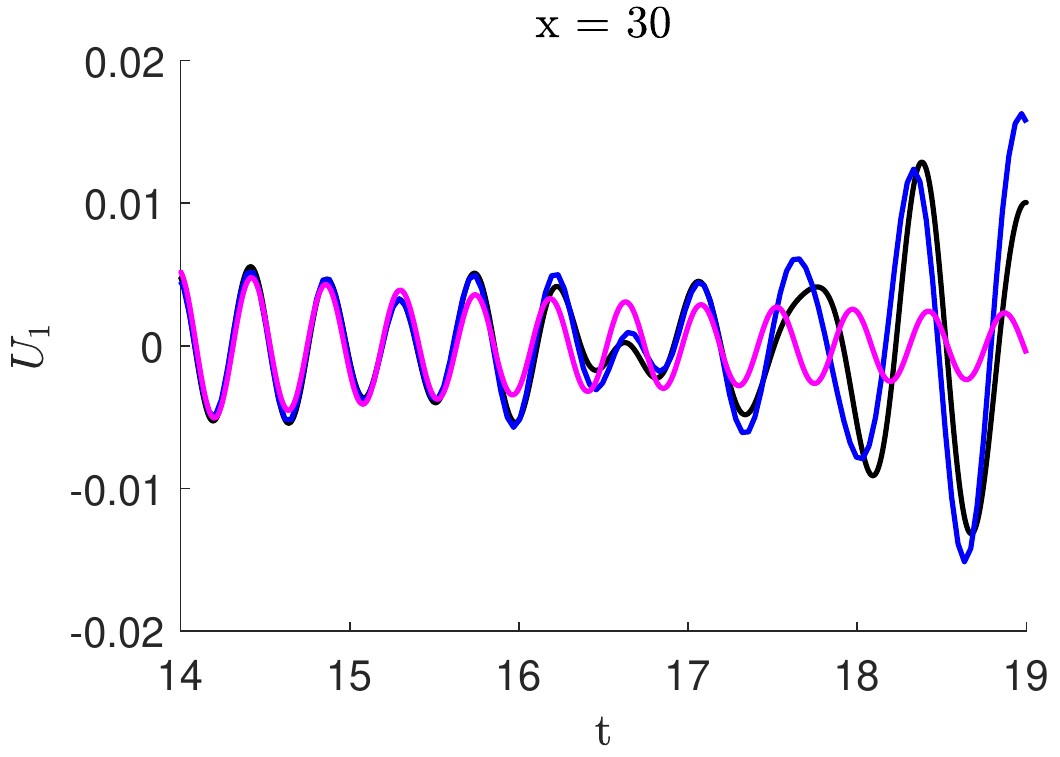}
\caption{Pulse response of the system (\ref{eqB310041}) -- (\ref{eqB310044}). Black: numerically calculated field using (\ref{e2205}), magenta: stationary phase method, blue: saddle point method. The right figure is a zoomed fragment of the left one}
\label{field_5D}
\end{figure}

Let us present results of the wavefield calculations using the stationary phase method and the saddle points method.  We set ${\gF} = (f, 0, 0, 0, 0)^T$, where $f$ is a Gaussian excitation:
\[
f=\sqrt{\frac{a}{\pi}}\exp(-at^2 ),\quad a = 25.
\]
 The results are shown in \figurename~\ref{field_5D}. The coordinate of the receiver is $x = 30$; the time dependence of the component $U_1$ is presented. The black line shows the results of numerical integration of (\ref{e2205}), the magenta line shows the stationary phase asymptotic, and the blue line show the saddle point method estimation. 
It follows from the figure that for times $14<t<19$ the saddle point method provides a 
considerably better result compared to the stationary phase method. What is more important, the saddle point method provides a physical signal that is completely ignored by stationary phase approximation. This signal is usually regarded as a forerunner \cite{Shanin2017}.

\section{Conclusion}
\label{sec7}

Transient processes in  waveguides described by the WaveFEM equation (\ref{e2101}) are studied. The field is 
represented as a sum of contour integrals (\ref{e2205}). It is shown that for large $x$ and $t$ this expression can be estimated with the help of the stationary phase method.  According to the method, the integral is estimated as a sum of terms (\ref{e2305}) provided by  real saddle points. On a numerical example (see \figurename~\ref{fig03}) we show that this approximation yields poor accuracy for some $x$ and $t$. This happens due to the presence of some other saddle points with non-zero imaginary part.  To take such points into account we continue the dispersion equation (\ref{e2204}) into the complex domain of variables $(k,\omega)$.  In the result a 2D surface $\gH$ in a 4D space is introduced. Then, the set of points at which the group velocity is real is found on $\gH$. The set is 1D and refereed as {\it the carcass of the dispersion diagram}. The points of the carcass are classified as {\it active} and {\it not active}. Active points can contribute to the field. 
We claim that the carcass can be used as a tool to study and classify the waveguides. For some waveguides, we build the carcasses numerically and study the pulses provided by the complex saddle points. Particularly, for the system (\ref{eqB310041}-\ref{eqB310044}) we show that pulse formed by the complex saddle points correspond to the forerunner.

%
%
%

\section*{Acknowledgements}
The study has been funded by RFBR, project number 19-29-06048.

\bibliography{biblio}
\bibliographystyle{ieeetr}

\end{document}